\documentclass[modern]{aastex631}

\received{June 24, 2022}
\revised{March 1, 2023}
\accepted{April 4, 2023}

\submitjournal{PSJ}

\shorttitle{Dual Band Radio Occultations of Venus}
\shortauthors{Akins et al.}

\graphicspath{{./}{figures/}}

\usepackage{amsmath}

\begin{document}

%
%



\title{Approaches for Retrieving Sulfur Species Abundances from Dual X/Ka Band Radio Occultations of Venus with EnVision and VERITAS}

\author{Alex B. Akins}
\affiliation{Jet Propulsion Laboratory, California Institute of Technology, Pasadena, California 91011, USA}

\author{Tatiana M. Bocanegra-Bahamón}
\affiliation{Jet Propulsion Laboratory, California Institute of Technology, Pasadena, California 91011, USA}

\author{Kuo-Nung Wang}
\affiliation{Jet Propulsion Laboratory, California Institute of Technology, Pasadena, California 91011, USA}

\author{Panagiotis Vergados} 
\affiliation{Jet Propulsion Laboratory, California Institute of Technology, Pasadena, California 91011, USA}

\author{Chi O. Ao}
\affiliation{Jet Propulsion Laboratory, California Institute of Technology, Pasadena, California 91011, USA}

\author{Sami W. Asmar}
\affiliation{Jet Propulsion Laboratory, California Institute of Technology, Pasadena, California 91011, USA}

\author{Robert A. Preston}
\affiliation{Jet Propulsion Laboratory, California Institute of Technology, Pasadena, California 91011, USA}

\begin{abstract}
The EnVision and VERITAS missions to Venus will fly with X and Ka band telecommunications channels which can be used to conduct radio occultation studies of Venus' atmosphere. While link attenuation measurements during prior S and X band occultation experiments have been used to determine vertical profiles of H$_2$SO$_4$ vapor abundance, the addition of the Ka band channel introduces greater sensitivity to the abundances of H$_2$SO$_4$ aerosols and SO$_2$ gas, permitting retrieval of their vertical profiles from dual band measurements. Such measurements would be valuable in the assessment of chemical and dynamical processes governing short and long-term variability in Venus' atmosphere. This paper considers the sensitivity of the X/Ka band radio attenuation measurement to these atmospheric constituents, as well as uncertainties and regularization approaches for conducting retrievals of these atmospheric sulfur species from future occultation experiments. We introduce methods for seeding maximum likelihood estimation retrievals using shape models and simple atmospheric transport constraints. From simulated retrievals, we obtain mean errors of the order of 0.5 ppm, 20 ppm, and 10 mg/m$^3$ for H$_2$SO$_4$ vapor, SO$_2$, and H$_2$SO$_4$ aerosol abundances, respectively, for simultaneous retrieval. 
\end{abstract}

\section{Introduction}
Spacecraft radio occultations (RO) have been used to accurately measure vertical profiles of the temperature and pressure of Venus' neutral atmosphere since the first use of the technique at Venus with Mariner V \citep{Fjeldbo1971}. Additionally, the observed excess attenuation of the radio link signal as it traverses the lower atmosphere has been used to infer the abundance of H$_2$SO$_4$ vapor \citep{Steffes1982, Kolodner1998a}. Prior analyses of neutral atmosphere RO soundings have assessed trends in cloud-level temperature and H$_2$SO$_4$ vapor abundances with altitude, latitude, and time \citep{Jenkins1991, Withers2020, Jenkins1994, Patzold2007, Tellmann2012, Oschlisniok2012, Oschlisniok2021, Imamura2017}. As a result, the average atmospheric structure above 40 kilometers (the average penetration depth of prior RO measurements) is now well known over a wide range of latitudes and local times \citep{Ando2020a}, and local variations, such as vertically-progagating gravity waves, have been well characterized. The recent analysis of over 800 Venus Express radio occultations by \citet{Oschlisniok2021} has provided thus far the most comprehensive assessment of the distribution of H$_2$SO$_4$ vapor with latitude and time. \citet{Oschlisniok2021} compared these results to a 2D transport model and found that the observed H$_2$SO$_4$ vapor distribution was recreated by driving meridional circulation with Hadley and polar cells. In this model, enhanced abundances of H$_2$SO$_4$ vapor are found at low latitudes and high latitudes, while mid-latitudes are relatively depleted. While the high latitude enhancement of H$_2$SO$_4$ vapor is driven mostly by sedimentation of cloud aerosols, the low latitude enhancement is highly circulation dependent and relies on supply of H$_2$SO$_4$ from the lower branch of the Hadley cell. By assuming that the vertical distribution of H$_2$SO$_4$ vapor follows the saturation vapor pressure curve and is thus negligible above a certain altitude ($>$ 50 km), \citet{Oschlisniok2021} also provided estimates of the sub-cloud abundance of SO$_2$. These estimates suggest that sub-cloud SO$_2$ abundance was greater in the polar regions than near mid-latitudes over the course of the Venus Express mission. These and other analyses of Venus RO measurements have yielded valuable results that will be useful for future dynamical and chemical modeling of the Venus atmosphere. \\

With the recent selection of several NASA and ESA missions to Venus, it is worthwhile to consider how the design of future RO experiments can provide new insight into the state of Venus' atmosphere. Of particular interest is the possibility of dual X (8.4 GHz, 3.5 cm) and Ka (32 GHz, 0.94 cm) band radio occultations of the neutral atmosphere. Of the recently selected missions, both EnVision and VERITAS will be capable of conducting dual X/Ka band RO experiments, although only EnVision has designated such experiments within its baseline objectives. The use of a Ka band link during RO measurements is of interest due to the increased atmospheric attenuation experienced by a 32 GHz signal, which may permit the retrieval of atmospheric neutral species beyond H$_2$SO$_4$ vapor. As discussed by \citet{Akins2020a}, the 32 GHz opacity of H$_2$SO$_4$ cloud aerosols and SO$_2$ gas in the cloud-level atmosphere is high enough to noticeably affect radio signal propagation. The prospects of success for retrievals of SO$_2$ gas or H$_2$SO$_4$ aerosols from dual X/Ka band occultations, however, have yet to be thoroughly considered. \\

If vertical profiling of either of these neutral species could be accomplished with RO measurements, the benefit to our understanding of Venus' atmosphere would be considerable. SO$_2$ is one of the most abundant trace species in Venus' atmosphere, but the processes that govern its vertical distribution remain unclear. SO$_2$ is thought to originate in the atmosphere from persistent volcanic outgassing \citep{Bullock2001}, and its photolysis in the mesosphere is a key mechanism in the formation of the H$_2$SO$_4$ clouds.  Above the clouds, order of magnitude variations in SO$_2$ abundance have been observed on relatively short timescales \citep{Marcq2013, Vandaele2017a}, which could possibly result from strong, episodic injection from the lower atmosphere driven by volcanism. \citep{Glaze1999, Airey2015}. Over longer timescales, observations have suggested a persistent steep decrease in SO$_2$ abundance within Venus' cloud layer between the troposphere and mesosphere. This decrease is difficult to reconcile with the results of atmospheric chemical models and requires either an inhibition of vertical transport, chemical depletion, or dissolution of SO$_2$ within the clouds via unexpected mechanisms \citep{Bierson2020, Rimmer2021}. RO measurements of the vertical distribution of SO$_2$ within the clouds could perhaps provide insight into this depletion/inhibition processes, and on shorter timescales, they could also be used to identify strong injections of SO$_2$ from the lower atmosphere. Gaps also exist in our knowledge of Venus' lower cloud structure and how it varies with latitude and time. While Venus Express observations provide strong constraints on the cloud-top altitude as a function of latitude, inferences of cloud-base altitude are far more ambiguous \citep{Barstow2012, Haus2013}. Beyond in situ results, most notably the Pioneer Venus LCPS measurements \citep{Knollenberg1980}, the latitudinal variation in lower cloud mass-loading is also weakly constrained by observations. Knowledge of the cloud mass and its contribution to cloud opacity is important in consideration of the radiative energy balance of Venus' atmosphere \citep{Limaye2018a} and in circulation modeling \citep{Sanchez-Lavega2017}. \\

In this paper, we investigate the accuracy with which H$_2$SO$_4$ and SO$_2$ abundances can be retrieved from dual X/Ka band radio occultations with upcoming spacecraft missions. In Section \ref{sec:link_atten}, we discuss contributing factors to link attenuation measured during an RO experiment and their associated uncertainties, including uncertainties for models of Venus' atmospheric opacity inferred from laboratory studies. In Section \ref{sec:retmet}, we illustrate the ill-posed nature of dual X/Ka band retrievals of sulfur species and introduce a regularization procedure to simultaneously retrieve H$_2$SO$_4$ and SO$_2$ profiles with greater vertical resolution than previously possible. We then apply these procedures to conduct simulated retrievals. We discuss the performance of these algorithms and their implications for actual retrievals in Section \ref{sec:discussion}, and we provide concluding remarks in Section \ref{sec:conclusion}. Overall, we argue that vertically resolved measurements of SO$_2$ and H$_2$SO$_4$ aerosols should be sufficiently accurate (20 ppm and 10 mg/m$^3$ for SO$_2$, and H$_2$SO$_4$ aerosol abundances, respectively) to both determine the mean atmospheric abundances of both species (as a function of latitude and altitude) and also identify strong perturbations from the mean. Our results are particularly encouraging for the possible detection of volcanic injection of SO$_2$ into the upper troposphere. 

\section{Radio Link Attenuation and Uncertainties} \label{sec:link_atten}

During a one-way spacecraft-to-Earth RO experiment, a spacecraft orbiting Venus transmits a radio carrier wave signal towards receivers on Earth, and the center frequency and amplitude of the received tone are modified via atmospheric refraction and attenuation. We consider here the 2-D spherically symmetric RO geometry for sounding a refractive neutral atmosphere described by Figure 1 of \cite{Eshleman1973}, where the coordinate system is defined by the vector between the centers of Earth and Venus and the orthogonal vector within the plane containing the Earth, Venus, and the spacecraft. In this coordinate system, simple relationships exist between the observed Doppler shift $f$ of the received RO signal from the center frequency $f_0$ (with electromagnetic wavelength $\lambda_0$), the spacecraft velocity $v_t$ in the orthogonal direction to the Earth-Venus vector, the complement angle $\gamma$ of the spacecraft elevation with respect to the Earth-Venus vector, the distance $R_s$ of the spacecraft from the center of Venus and the occultation ray impact parameter $a$ and bending angle $\delta$. 

\begin{equation} \label{eq:fres}
f = (v_t / \lambda_0) \sin{\delta}
\end{equation}

\begin{equation} 
a = R_s \cos{(\gamma - \delta)}
\end{equation}

The refractive index $n$ can be determined directly from knowledge of the ray impact parameter and bending angle through an inverse Abel transform. 

\begin{equation} \label{eq:refractderiv}
\ln{n(a)} = \frac{1}{\pi} \int_{a}^\infty \frac{\delta da'}{\sqrt{a'^2-a^2}} 
\end{equation}

Assuming perfect antenna pointing, the Doppler-shifted RO signal is also attenuated via refractive defocusing and neutral atmosphere gas absorption. The refractive defocusing contribution $L$ is frequency-indepdendent and defined below by the experiment geometry under the assumption that the Earth is significantly farther from the spacecraft than Venus \citep{Eshleman1973, Oschlisniok2012}. 

\begin{equation} 
L = -10 \log_{10}\left(\Phi_1 \Phi_2\right)
\end{equation}

\begin{equation*}
\Phi_1 = \left(\sec{\delta} - \frac{D}{a}\tan{\delta}\right)^{-1}
\end{equation*}

\begin{equation*} 
\Phi_2 = \left(1 + \left(a \tan{\delta} - D \sec{\delta}\right) \frac{d\delta}{da}\right)^{-1}
\end{equation*}

\begin{equation*} 
D = R_s \left[\sin{(\gamma - \delta)} + \cos{(\gamma - \delta)}\tan{\frac{\delta}{2}}\right]
\end{equation*}

Once the contribution to total link attenuation from refractive defocusing is subtracted, the resulting excess attenuation $\tau$ can be converted to absorptivity $\alpha$ profiles in dB/km units via an inverse Abel transform, which is written in terms of attenuation, absorptivity, ray impact parameter, and ray periapse altitude $r$ \citep{Jenkins1991, Oschlisniok2012}

\begin{equation} \label{eq:alphaderiv}
\alpha(a) = - \frac{da}{dr} \frac{1}{\pi a} \frac{dF}{da}
\end{equation}

\begin{equation*}
    F = \int_{a}^\infty \frac{\tau a' da'}{\sqrt{a'^2 - a^2}}
\end{equation*}

The resulting absorptivity profile can then be used with temperature and pressure profiles derived from the measured refractivity to retrieve the abundance of atmospheric absorbers. 

\subsection{Random uncertainty in absorptivity profiles}  
Knowledge of random uncertainties in the measurement of the radio link Doppler shift and signal strength can be used to determine uncertainties in the resulting absorptivity profiles. In this section, we review the calculation of uncertainties in RO-inferred absorptivity based on the discussions in \cite{Lipa1979, Jenkins1991, Oschlisniok2012} and state our assumptions regarding Doppler shift and signal strength statistics for X and Ka band RO measurements which are relevant to our simulated retrievals of atmospheric composition. 

In the reconstruction of the received carrier tone at the ground station, different sources of errors contribute to the uncertainties of the frequency and power estimates: instrumental (onboard the spacecraft and at the receiving system) and propagation random errors (introduced by the presence of interplanetary plasma, Earth's ionosphere and troposphere). Neglecting errors in the trajectory of the spacecraft, the variance of the frequency time series estimates is given by the summation of the contributions of thermal and phase noise
\begin{equation}
\sigma^2_{f} = \frac{2B N_0/C}{(2\pi \tau)^2} +\sigma^2_{AD}f_0^{2}
\label{eq:sigmaf}
\end{equation}
where $B=1$ Hz is the noise bandwidth, $N_0$ is the noise power density, $C$ is the signal power and $\tau$ is the integration time, $\sigma_{AD}$ is the Allan deviation of the phase noise and $f$ is the nominal signal frequency. We assume the phase noise is dominated by the onboard frequency standard (assuming $ \sigma_{AD} \sim 5\times 10^{-13}$ at $\tau=0.1$\,s, \citet{Hausler2006a}) and it is constant throughout the occultation event. The $C/N_0$ ratio will decrease during the occultation as the signal probes deeper in the atmosphere, increasing the uncertainty in the frequency measurement through Equation \ref{eq:sigmaf}. The $C/N_0$ ratio also describes the noise in the received signal power $p$ in dB-Hz (for a signal with linear amplitude $s_a$). The relationships between received signal power $p$, signal amplitude $s_a$ and their corresponding uncertainties $\sigma_p$ and $\sigma_a$ are given in Equation \ref{eq:sigpower}.

\begin{equation} \label{eq:sigpower}
p = 10 \log_{10}s_a^2
\end{equation}

\begin{equation*} 
\sigma_a = \left(\sqrt{10^{\frac{C/N_0}{10}}}\right)^{-1}
\end{equation*}

\begin{equation*}
\sigma_p = \frac{\partial p}{\partial s_a} \sigma_a 
\end{equation*}

We assume a top-of-atmosphere $C/N_0$ ratio of 70 dB for X band and 80 dB Ka band, which is consistent with the notional design for the EnVision radio science experiment \citep{Envision2021}. We also increase the phase noise contribution by adding a constant value (0.2 Hz at X band, 0.8 Hz at Ka band) to the Doppler shift uncertainty, which is consistent with the uncertainties observed for the Magellan orbit 3212 X band occultation.

To determine estimated of uncertainty for the simulated RO retrievals discussed in this paper, we employ a forward and inverse RO simulator using Equations \ref{eq:fres}-\ref{eq:alphaderiv} (similar to \citet{Jenkins1992}). For the model atmospheric compositions discussed in later sections, simulated power and frequency time series are derived for an occultation experiment at X and Ka band by a spacecraft in a circular orbit at 250 km altitude. From the values of absorptivity, refractive index and impact parameter ($a=nr$) corresponding to the models, bending angle and signal attenuation are derived using forward Abel transforms

\begin{equation}
  \delta(a) = -2a \int^{\infty}_{r}\frac{dn}{da(r')}\frac{dr'}{\sqrt{a(r')^2 - a(r)^2}}
\end{equation}

\begin{equation}
\tau(r) = 2 \int^{\infty}_{r}\frac{\alpha(r) a(r') dr'}{\sqrt{a(r')^2 - a(r)}}
\end{equation}

The uncertainty in the inferred absorptivity from the RO measurement can be determined via linear propagation of errors \citep{Jenkins1991}, starting from the frequency and power uncertainties $\sigma_f$ and $\sigma_p$. We assume that there is no covariance between the recorded signal power and Doppler shift at different times, which is appropriate for the assumed time and bandwidth integrations (see discussion in \cite{Lipa1979}). In the simplified RO geometry, the corresponding covariance matrices for the impact parameter $C_a$ and bending angle $C_\delta$ are diagonal and computed as $C^{i=j}_a = \left(\frac{\partial a}{\partial f} \sigma_f\right)^2$ and $C^{i=j}_\delta = \left(\frac{\partial \delta}{\partial f} \sigma_f\right)^2$, where

\begin{equation}
\frac{\partial \delta}{\partial f} = \frac{\lambda_0 / v_t}{\sqrt{1 - (\lambda_0 / v_t)^2 f^2}}
\end{equation}

\begin{equation} 
\frac{\partial a}{\partial f} = R_s \sin{(\gamma - \delta)} \frac{\partial \delta}{\partial f} 
\end{equation}

In this convention, the $i$ index corresponds to the matrix row and the $j$ index corresponds to the column. Off-diagonal terms are introduced into the covariance matrices for terms that are the result of an inverse Abel transform, such as the inferred refractive index and absorptivity. In \cite{Lipa1979} Appendix A, a procedure is provided for the determination of the refractive index covariance matrix using a midpoint-rule Riemann sum discretization of an alternate form of Equation \ref{eq:refractderiv}. 

\begin{equation}
\ln{n(a)} = \frac{1}{\pi} \int_{\delta(a)}^\infty \ln\left[\frac{a'}{a} + \sqrt{(a'/a)^2 - 1}\right] d\delta(a') \approx \frac{1}{\pi} \sum_{j=1}^n h_{ij} (\delta_{j + 1} - \delta_j)
\end{equation}

\begin{equation*} 
h_{ij} = \log\left[\frac{a_{j + 1} + a_j}{2a_i} + \sqrt{\left(\frac{a_{j + 1} + a_j}{2a_i}\right)^2 - 1}\right]
\end{equation*}

The covariance matrix $C_n$ can then be determined as $C_n=T_{na} C_a T_{na}^T + T_{n\delta} C_\delta T_{n\delta }^T$, where the $T_{na}$ and $T_{n\delta}$ matrices are lower triangular and determined as 

\begin{equation} 
T^{i,j < i}_{na} = \frac{\partial n_i}{\partial a_j} = \frac{\partial h_{ij}}{\partial a_j}(\delta_{j + 1} - \delta_j) + \frac{\partial h_{i, j-1}}{\partial a_j}(\delta_{j} - \delta_{j-1})
\end{equation}

\begin{equation*}
\frac{\partial h_{ij}}{\partial a_j} = \frac{\partial h_{i, j-1}}{\partial a_j} = \left(\sqrt{\left(a_j - 2 a_i + a_{j+1}\right)\left(a_j + 2 a_i + a_{j+1}\right)}\right)^{-1}
\end{equation*}

\begin{equation} 
T^{i,j < i}_{n\delta} = \frac{\partial n_i}{\partial \delta_j} = h_{i, j - 1} - h_{ij}
\end{equation}

The uncertainty in the inferred excess signal attenuation $\tau$ is the sum of the uncertainty in the signal power measurement and the uncertainty in the refractive defocusing estimate as $C^{i=j}_\tau = \sigma_p^2 + \left(\frac{\partial L}{\partial a}\right)^2 C_a + \left(\frac{\partial L}{\partial \delta}\right)^2 C_\delta$, where 

\begin{equation} 
\frac{\partial L}{\partial a} = \frac{10}{\ln{10}} \frac{1}{\Phi_1 \Phi_2}\left[\frac{\partial \Phi_1}{\partial a} \Phi_2 + \frac{\partial \Phi_2}{\partial a} \Phi_1 \right]
\end{equation}

\begin{equation*} 
\frac{\partial L}{\partial \delta} = \frac{10}{\ln{10}} \frac{1}{\Phi_1 \Phi_2}\left[\frac{\partial \Phi_1}{\partial \delta} \Phi_2 + \frac{\partial \Phi_2}{\partial \delta} \Phi_1 \right]
\end{equation*}

\begin{equation*}
\frac{\partial \Phi_1}{\partial a} = -\frac{D \tan{\delta}}{(a\sec{\delta} - D \tan{\delta})^2}
\end{equation*}

\begin{equation*}
\frac{\partial \Phi_1}{\partial \delta} = \frac{a\sec{\delta}(D\sec{\delta}-a\tan{\delta})}{(a\sec{\delta} - D \tan{\delta})^2}
\end{equation*}

\begin{equation*}
\frac{\partial \Phi_2}{\partial a} = -\frac{\frac{\partial \delta}{\partial a} \tan{\delta}}{\left[1 + \left(a \tan{\delta} - D\sec{\delta}\right) \frac{\partial \delta}{\partial a} \right]^2}
\end{equation*}

\begin{equation*}
\frac{\partial \Phi_2}{\partial \delta} = -\frac{\frac{\partial \delta}{\partial a} \left(\sec{\delta} \left(a \sec{\delta} - D \tan{\delta}\right) \right)}{\left[1 + \left(a \tan{\delta} - D\sec{\delta}\right) \frac{\partial \delta}{\partial a} \right]^2}
\end{equation*}

Next, the covariance matrix of the intermediate inverse Abel transform term in Equation \ref{eq:alphaderiv} is determined as as $C_F = T_{Fa}C_aT_{Fa}^T + T_{F\tau}C_aT_{F\tau}^T$. The lower diagonal $T_{Fa}$ and $T_{F\tau}$ matrices are computed using a similar discretization to that employed in the calculation of the $T_n$ matrices. 

\begin{equation} 
F = \sum_{j=1}^n g_{ij} \frac{\tau_{j} + \tau_{j + 1}}{2} 
\end{equation} 

\begin{equation*} 
g_{ij} =\frac{ \frac{a_{j} + a_{j + 1}}{2} \left(a_{j + 1} - a_j\right)}{\sqrt{\left(\frac{a_j + a_{j + 1}}{2}\right)^2 - a_0^2}}
\end{equation*} 

\begin{equation} 
T^{i, j<i}_{F\tau} = \frac{\partial F}{\partial \tau_j} = \frac{1}{2}(g_{ij} + g_{i, j-1})
\end{equation}

\begin{equation*} 
T^{i, j<i}_{Fa} = \frac{\partial F}{\partial a_j} = \frac{\partial g_{ij}}{\partial a_j}(\frac{\tau_{j + 1} + \tau_j}{2}) + \frac{\partial g_{i, j-1}}{\partial a_j}(\frac{\tau_{j} + \tau_{j-1}}{2})
\end{equation*}

\begin{equation*} 
\frac{\partial g_{ij}}{\partial a_j} = -\frac{\partial g_{i,j-1}}{\partial a_j} = -\frac{(a_j + a_{j + 1})^3 - 8a_i^2a_j}{((a_j + a_{j + 1})^2 - 4a_i^2)^{3/2}}
\end{equation*}

Finally, the absorptivity covariance is computed as $C_\alpha = \left(\frac{\partial \alpha}{\partial F}\right)^2 C_F + \left(\frac{\partial \alpha}{\partial n}\right)^2 C_n + \left(\frac{\partial \alpha}{\partial a}\right)^2 C_a$, where 

\begin{equation} 
\frac{\partial \alpha}{\partial a} = \frac{dF}{dr} \frac{1}{\pi a^2}, \quad \frac{\partial \alpha}{\partial n} = -\frac{dF}{da} \frac{1}{\pi a}, \quad \frac{\partial \alpha}{\partial F} = -\frac{1}{\Delta r \pi a}
\end{equation}

\begin{figure}
    \centering
    \includegraphics[width=\textwidth]{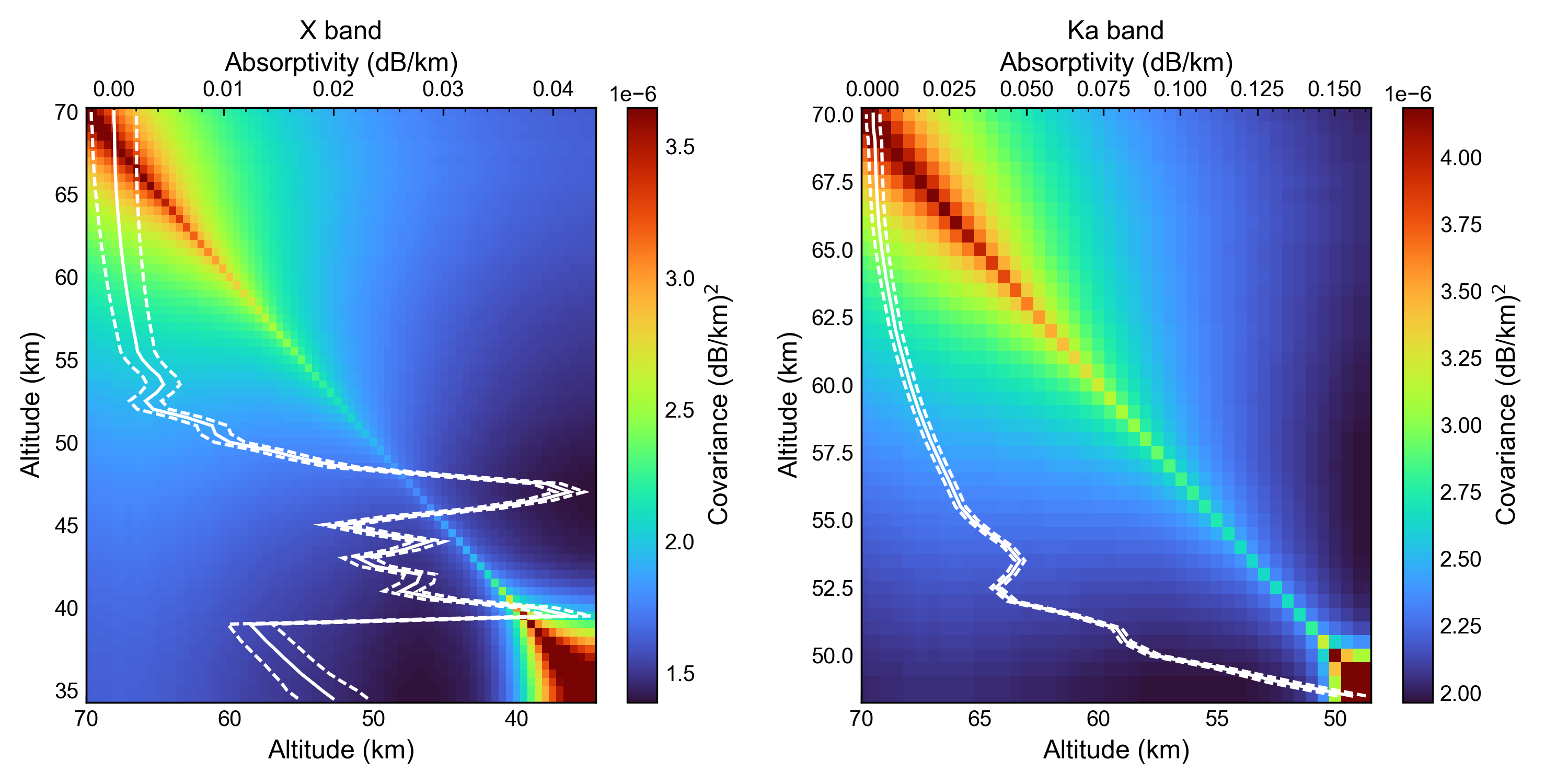}
    \caption{Covariance matrices computed for X and Ka band atmospheric absorptivity from RO simulations for the Set 1 model atmosphere. The corresponding 1$\sigma$ uncertainties are shown in context bracketing the absorptivity profiles.}
    \label{fig:abs_covar}
\end{figure}

Examples of these covariance matrices computed at in uniform 0.5 km intervals for X and Ka band occultations are shown in Figure \ref{fig:abs_covar} for the Set 1 model atmosphere in Table \ref{tab:datset}. In addition to the offsets introduced to the Doppler shift uncertainty, the covariance profiles are also multiplied by a factor of 15. This factor was determined by comparing the results of simulations with $C/N_0=60$ dB to the reported results of Magellan occultations. \cite{Jenkins1994} applied a similar correction factor in their presentation of the Magellan RO results, and this was intended as compensation for small-scale fluctuations in the signal power.  For both bands, the diagonal terms are stronger than the off-diagonal terms due to the form of the denominator in the inverse Abel transform expressions. For both bands, the uncertainties are greater at higher altitudes and lower altitudes due to lower rate of sampling and proximity to the signal attenuation limit, respectively. The signal attenuation limits (i.e. deepest sounding depth) range between 35-40 km for X band and 45-50 km for Ka band.

\subsection{Neutral atmosphere absorptivity} \label{sec:opac_models}
The consequential microwave absorbers at X and Ka band in the atmosphere of Venus are the bulk CO$_2$/N$_2$ atmosphere, H$_2$SO$_4$ vapor and aerosols, and SO$_2$. Continuum and spectral line models of the microwave opacity of these species in the atmosphere of Venus have been derived from laboratory measurements under simulated Venus conditions \citep{Ho1966, Fahd1991, Fahd1992, Kolodner1998a, Akins2020a}. While H$_2$SO$_4$ vapor and SO$_2$ opacity are described by spectral line models, single frequency expressions at X and Ka band have been derived as linear functions of gas volume mole fraction, which are reviewed in this section. Additionally, first order uncertainties have been derived for opacity model parameters based on fits to respective laboratory data sets. These uncertainties were determined in the Bayesian sense, where for some dataset $\mathbf{x}$ with uncertainty $\mathbf{\sigma}$ and model parameter $a$, the probability $P(a|\mathbf{x})$ can be estimated using Equation \ref{eq:stats}. The resulting Gaussian-like probability distribution can be used to determine 2$\sigma$ uncertainties for each model parameter. Since the covariance of the model parameters are not considered here, these are conservative 2$\sigma$ estimates.

\begin{equation} \label{eq:stats}
P(a | \mathbf{x}) \propto \prod_{i=1}^n P(x_i|a, \sigma_i)
\end{equation}

For H$_2$SO$_4$ vapor, \citet{Kolodner1998a} and \citet{Akins2020a} determined single frequency S, X, and Ka Band models for H$_2$SO$_4$ vapor opacity $\alpha$ as a function of pressure $p$ in atmospheres, temperature $T$ in Kelvins, and volume mole fraction $q$. The temperature dependence is given by the $\theta$ term, where $\theta=553/T$ for the H$_2$SO$_4$ vapor model. The values of these parameters and their derived 2$\sigma$ uncertainties are shown in Table \ref{tab:h2so4_pars}

\begin{equation} \label{eq:cont_model}
\alpha = a_1 p^{a_2}\theta^{a_3} q \quad \mbox{dB/km}
\end{equation}

\begin{table}[]
\caption{H$_2$SO$_4$ vapor opacity model parameters and 2$\sigma$ uncertainties}
\begin{center}
\begin{tabular}{cccc}
\hline
Band & $a_1$ & $a_2$ & $a_3$  \\ \hline
S (2.26 GHz) & 106.58 $\pm$ 2.90 & 1.333 $\pm$ 0.02 & 3.2 $\pm$ 0.2 \\
X (8.39 GHz) & 451.76 $\pm$ 3.04 & 1.283 $\pm$ 0.005 & 3.0 $\pm$ 0.2 \\
Ka (32 GHz) & 2586.66 $\pm$ 421.64 & 1.092 $\pm$ 0.12 & 3.0 $\pm$ 0.2 \\ \hline
\end{tabular}
\end{center}
\label{tab:h2so4_pars}
\end{table}

For H$_2$SO$_4$ aerosol, \citet{Fahd1991} determined parameters for a Cole and Cole model of the complex dielectric constant $\epsilon_r=\epsilon_r' - j\epsilon_r''$. This model is expressed in terms of a static dielectric constant $\epsilon_{rs}$, a high-frequency dielectric constant $\epsilon_{r\infty}$, and relaxation constants $\tau$ and $a$. This model is used to determine absorption at a given wavelength $\lambda$ for an aerosol mass with bulk density M (in units of mg of aerosol per atmosphere volume in m$^3$) and a characteristic solution liquid density of $\rho$ (mg of liquid per liquid volume in m$^3$). We assume that the aerosol particle diameter is small enough (10s of microns) such that scattering does not need to be considered, which is an acceptable assumption for Venus' atmosphere \citep{Fahd1991}.

\begin{equation} 
\epsilon_r = \epsilon_{r \infty} + \frac{\epsilon_{rs} - \epsilon_{r \infty}}{1 + (j \omega \tau)^{1 - a}}
\end{equation}

\begin{equation} 
\alpha = \frac{246M\epsilon_r''}{\rho \lambda\left[\left(\epsilon_r' + 2\right)^2 + \left(\epsilon_r''\right)^{2}\right]} \quad \mbox{dB/km}
\end{equation}

\citet{Fahd1991} made measurements of 85\% and 99\% H$_2$SO$_4$ solutions, which resulted in different models. The dielectric model parameters and their 2$\sigma$ uncertainties are shown in Table \ref{tab:cloud_pars}. These models were fit to Fahd's Ka and W band measurements, and while limiting the fit to only Ka Band changes the model parameters, the broadband fit is preferable due to the possible presence of systematic offsets. Uncertainty in cloud weight percent H$_2$SO$_4$, which may range from 75\% to 99 \% in the atmosphere of Venus, also contributes to retrieval uncertainties.

\begin{table}[]
\caption{H$_2$SO$_4$ aerosol dielectric model parameters and 2$\sigma$ uncertainties}
\begin{center}
\begin{tabular}{cccc}
\hline
H$_2$SO$_4$ Weight Percent & $\epsilon_{r \infty}$ & $\tau$ & $a$ \\ \hline
85 \% & 3.393 $\pm$ 0.290 & (1.78 $\pm$ 0.02) $\times 10^{-11}$ & 0.113 $\pm$ 4.6e-3  \\
99 \% & 2.319 $\pm$ 0.065 & (2.576 $\pm$ 0.04) $\times 10^{-10}$ & 0.390 $\pm$ 2.0e-3  \\ \hline

\end{tabular}
\end{center}
\label{tab:cloud_pars}
\end{table}

\citet{Fahd1992} also determined a spectral line model for SO$_2$ opacity which has been corroborated over a range of frequencies and temperatures \citep{Suleiman1996, Bellotti2015, Steffes2015}. The absorption of gaseous SO$_2$ can be expressed as the product of the line center absorption and a Van Vleck-Weisskopf lineshape function. 

\begin{equation}
    \alpha = A_{max}F_{VVW}(\nu, \Delta \nu) \quad \mbox{dB/km}
\end{equation}

\begin{equation} \label{deltanu}
\Delta\nu = \gamma p\left(\frac{T_o}{T}\right)^{n} \quad \mbox{MHz}
\end{equation}

The free parameters for the model are the linewidth parameters $\gamma$ and their temperature dependence $n$ for SO$_2$-SO$_2$ and SO$_2$-CO$_2$ broadening, which are shown in Table \ref{tab:so2_pars} with their 2$\sigma$ uncertainties. Single frequency expressions at S, X, and Ka Band and their uncertainties have  also been derived by fitting a model with the form of Equation \ref{eq:cont_model} (and $\theta = 300/T$) to the spectral line model predictions. Due to the nonlinear relationship between total mixture pressure in the atmosphere and SO$_2$ opacity, separate expressions are given that are applicable below and above 1.5 atmosphere mixture pressure, respectively. Note that this is not necessary for H$_2$SO$_4$ vapor, which is largely depleted above the 1 atmosphere altitude. The resulting parameters are shown in Table \ref{tab:so2_cont_pars}.

\begin{table}[]
\caption{SO$_2$ spectral line model parameters and 2$\sigma$ uncertainties}
\begin{center}
\begin{tabular}{ccc}
\hline
Broadening Gas & $\gamma$ (MHz/torr) & $n$ \\ \hline
SO$_2$ & 16 $\pm$ 1.58 & 0.85 $\pm$ 0.11 \\
CO$_2$ & 7 $\pm$ 0.91  & 0.85 $\pm$ 0.07 \\ \hline
\end{tabular}
\end{center}
\label{tab:so2_pars}
\end{table}

\begin{table}[]
\caption{SO$_2$ continuum model parameters and 2$\sigma$ uncertainties}
\begin{center}
\begin{tabular}{ccccc}
\hline
Band & $p$ & $a_1$ & $a_2$ & $a_3$  \\ \hline
S (2.26 GHz) & $<$ 1.5 atm & 1.36 $\pm$ 0.03 & 1.19 $\pm$ 0.02 & 2.65 $\pm$ 0.01 \\
& $\geq$ 1.5 atm & 1.36 $\pm$ 0.02 & 1.18 $\pm$ 0.02 & 2.74 $\pm$ 0.03 \\
X (8.39 GHz) & $<$ 1.5 atm & 21.68 $\pm$ 0.20 & 0.89 $\pm$ 0.04 & 2.48 $\pm$ 0.02 \\
& $\geq$ 1.5 atm & 19.10 $\pm$ 0.20 & 1.15 $\pm$ 0.03 & 2.72 $\pm$ 0.04 \\
Ka (32 GHz) & $<$ 1.5 atm & 309.10 $\pm$ 4.50 & 1.079 $\pm$ 0.003 & 2.66 $\pm$ 0.02 \\ 
& $\geq$ 1.5 atm & 288.94 $\pm$ 3.60 & 1.15 $\pm$ 0.01 & 2.75 $\pm$ 0.03 \\ \hline
\end{tabular}
\end{center}
\label{tab:so2_cont_pars}
\end{table}

\citet{Ho1966} made measurements of CO$_2$ opacity at 9 GHz and determined a model as a function of frequency $\nu$ in GHz, temperature $T$ in Kelvins, and pressure $p$ in atmospheres that was confirmed by \citet{Steffes2015}. The values of these parameters and their derived 2$\sigma$ uncertainties from the data of \citet{Steffes2015} are shown in Table \ref{tab:co2_pars}

\begin{equation} 
\alpha = a_1 p^{a_2}T^{a_3}\nu^{a_4} q \quad \mbox{dB/km}
\end{equation}

\begin{table}[]
\caption{CO$_2$ opacity model parameters and 2$\sigma$ uncertainties}
\begin{center}
\begin{tabular}{cccc}
\hline
$a_1$ & $a_2$ & $a_3$ & $a_4$ \\ \hline
(1.15 $\pm$ 0.06) $\times 10^{8}$ & 2 $\pm$ 0.01 & -5 $\pm$ 0.05 & 2 $\pm$ 0.05 \\ \hline
 \end{tabular}
\end{center}
\label{tab:co2_pars}
\end{table}

\begin{figure}[t]
\begin{center}
\begin{minipage}[c]{0.48\textwidth}
    \centering
    \includegraphics[width=\textwidth]{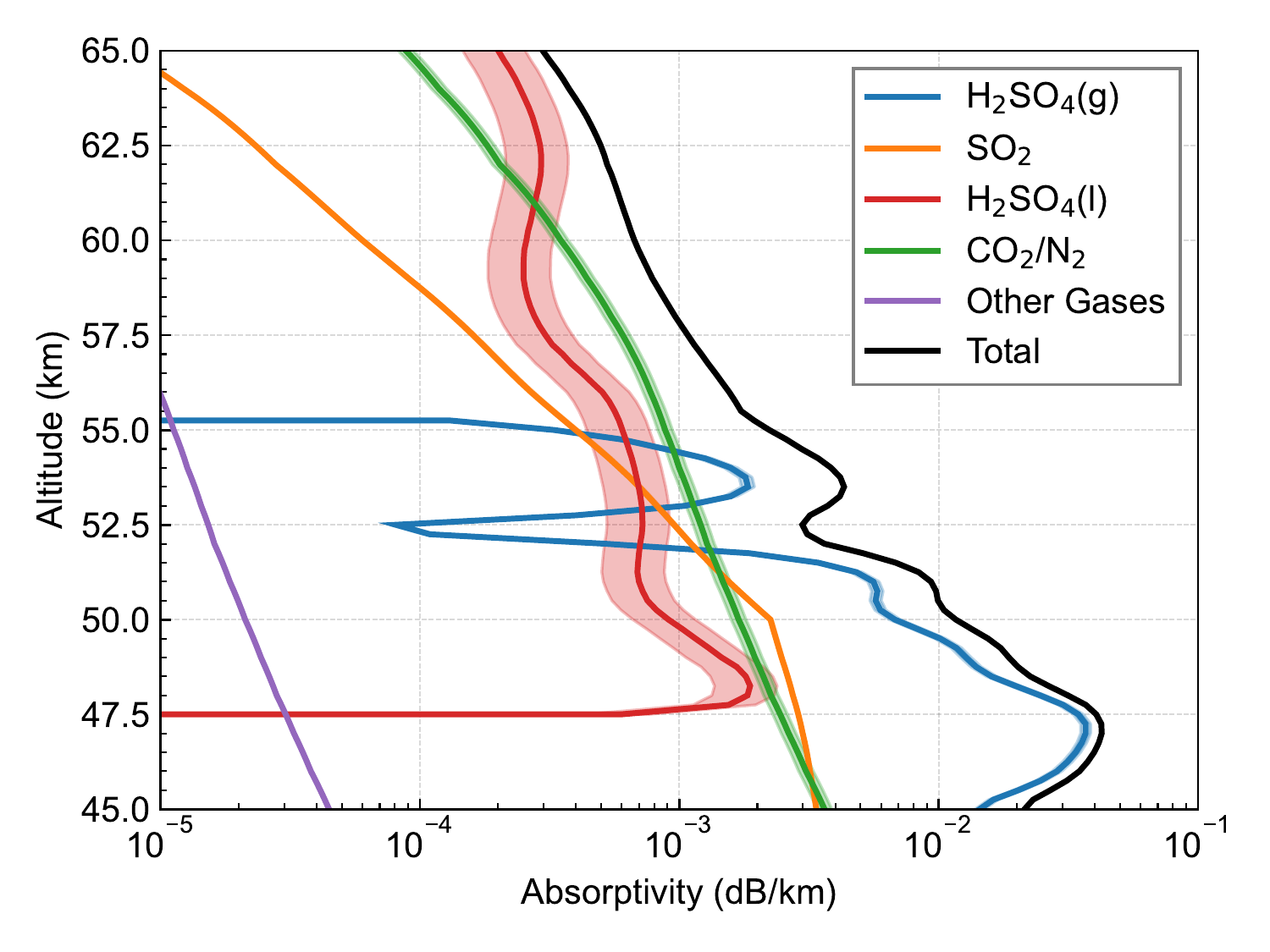}
\end{minipage}
\begin{minipage}[c]{0.48\textwidth}
    \centering
    \includegraphics[width=\textwidth]{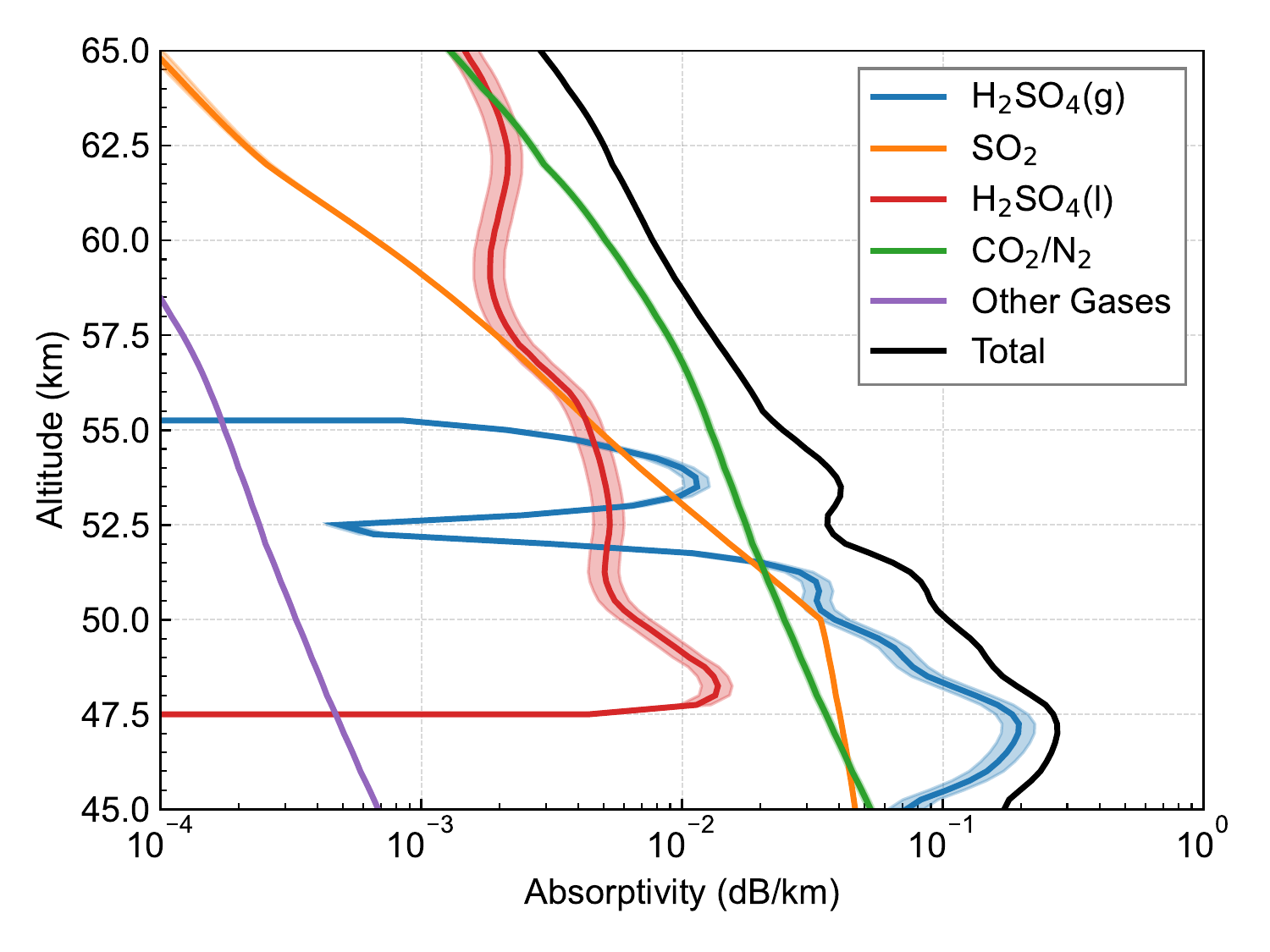}
\end{minipage}
\caption{X (left) and Ka (right) band atmospheric opacity of neutral atmosphere microwave absorbers for an equatorial model atmosphere. The bulk CO$_2$/N$_2$ atmosphere and sulfur species dictate the microwave opacity of the atmosphere, and the contribution of other trace gases is minimal. Shaded regions show 1$\sigma$ uncertainties associated with opacity models.}
\label{fig:gas_opac}
\end{center}
\end{figure}

These uncertainties can be then converted to uncertainties in retrieved gas abundances from RO measurements via standard propagation of errors methods \citep{Oschlisniok2012}. Figure \ref{fig:gas_opac} shows an example of X and Ka Band atmospheric absorptivity and 1$\sigma$ uncertainties (1$\sigma$ has been convention for Venus RO absorptivity measurements) for a model atmosphere (see Section \ref{sec:mod_atm}) of these constituents. The largest sources of uncertainty at Ka band are associated with H$_2$SO$_4$ vapor and aerosol, both of which exhibit a 1$\sigma$ uncertainty near 13\%. While the H$_2$SO$_4$ vapor uncertainty is a result of the laboratory measurement uncertainties, the aerosol uncertainty is almost entirely due to uncertainty in the weight percent H$_2$SO$_4$ of the aerosols themselves. Uncertainties are shown assuming a range of H$_2$SO$_4$ aerosol weight percents between 85\%-99\%. This uncertainty can be somewhat reduced if a reasonable vertical profile of H$_2$SO$_4$ weight percent can be assumed, such as that of \citet{Krasnopolsky2015}. Also included in Figure \ref{fig:gas_opac} is the contribution from other gases whose volume mole fraction exceeds 1 ppm in this altitude range, specifically CO, OCS, and H$_2$O. At its highest, the contribution of these additional trace gases to atmospheric opacity is near 1.5\%, and uncertainties in their abundances contribute less than 1\% to the overall uncertainty in Ka band RO measurements. From Figure \ref{fig:gas_opac}, the relatively increased contribution at Ka band of SO$_2$ and H$_2$SO$_4$ aerosol to the total absorptivity profile measured during an RO experiment is apparent, hence the interest in dual X/Ka band RO for constraining the vertical distribution of these species.

\section{Simulated Retrievals} \label{sec:retmet}
With the relevant uncertainties established, we can now consider approaches to retrieve abundance profiles of H$_2$SO$_4$ and SO$_2$ from dual band RO measurements. These approaches are tested via simulated retrievals where several models of Venus' atmosphere are used as ground truth. We assume that raw measurements of X and Ka band amplitude and tone frequency have been converted to atmospheric refractivity and absorptivity profiles using geometric optics methods and that the associated uncertainties have been established following the discussion in the previous section. Since we are interested in the intrinsic retrieval accuracy, we consider only random uncertainties in these simulated retrievals and not the systematic uncertainties in opacity models discussed in the previous section. Our simulated retrievals span a range from 70 km altitude, where Ka band absorptivity will first be measurable, to the Ka Band signal attenuation limit for each model atmosphere.

\subsection{The Ill-Posed Retrieval Problem} \label{sec:ill-pose}
Since H$_2$SO$_4$ vapor, aerosol and SO$_2$ contribute non-negligibly to X and Ka band link attenuation and the frequency dependences of H$_2$SO$_4$ aerosol and SO$_2$ are similar, the retrieval of their respective abundances is an under-determined ill-posed problem. Since H$_2$SO$_4$ vapor is the strongest absorber, its abundance can be determined with accuracy comparable to that of dual S/X band RO retrievals \citep{Jenkins1994}. The ill-posed nature of retrieving sulfur species abundances beyond H$_2$SO$_4$ vapor can be illustrated by assessing the uniqueness of retrievals at a single altitude. It is assumed that the atmospheric temperature and pressure are known ($T=350$ K, $P=1$ bar), as well as the abundance of H$_2$SO$_4$ vapor (10 ppm). The relationship between atmospheric abundances of trace species $\mathbf{x} = \left[q_{H_2SO_4(g)}, M_{H_2SO_4(l)}, q_{SO_2}\right]$ and measured absorptivity $\mathbf{y} = \left[y_{X}, y_{Ka}\right]$ is established via a forward model $\mathbf{y} = \mathbf{Kx}$. The values of the $\mathbf{K}$ matrix are the derivatives of the linear opacity expressions for each absorber (see Section \ref{sec:opac_models}) with respect to abundance (volume mole fraction $q$ or bulk density $M$). A 10\% uncertainty (consistent with the expected uncertainty range, see Figure \ref{fig:abs_covar}) in the measurement of $\mathbf{y}$ is included and represented as the diagonal matrix $\mathbf{S_{y}}$. The probability of a particular atmospheric composition from a given measurement $P(x|y)$ can then be determined following \citet{Rodgers2000}. 

\begin{equation} \label{eqn:lgprb}
-2\mbox{ln}P(x|y) \propto \mathbf{(y - Kx)^TS_y^{-1}(y-Kx)}
\end{equation}

\begin{figure}[t]
\begin{center}
\includegraphics[width=0.75\textwidth]{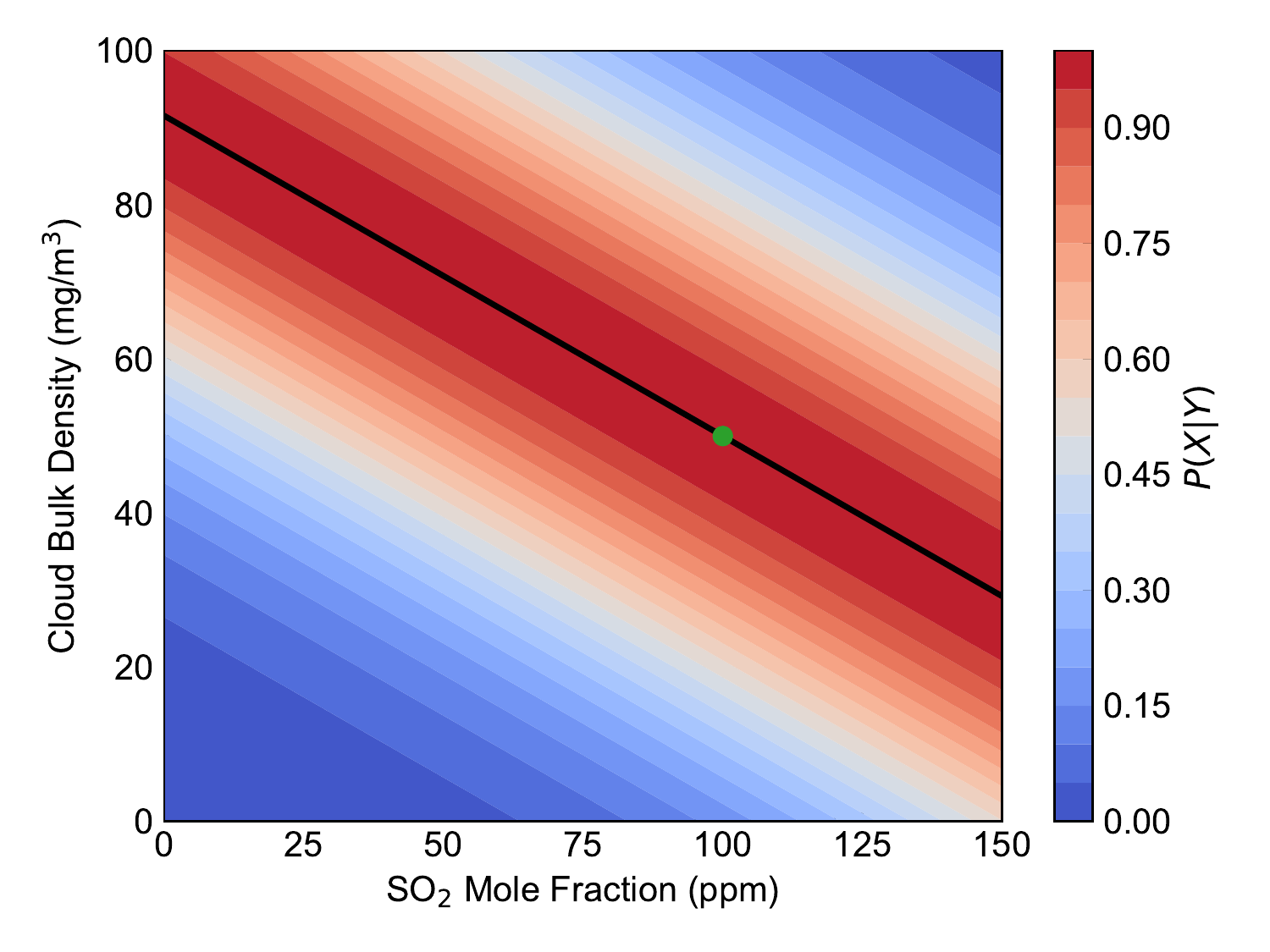}
\caption{Probability of SO$_2$ and H$_2$SO$_4$ aerosol abundance combinations consistent with simulated radio link absorption at X and Ka Band at an altitude of 50 km with an H$_2$SO$_4$ volume mole fraction of 10 ppm. The true abundances used for the simulations are shown as a green dot, and a line is shown representing the range of solutions possible for an error-free measurement.}
\label{fig:degen}
\end{center}
\end{figure}

Figure \ref{fig:degen} shows the resulting probability distribution for a given abundance of H$_2$SO$_4$ aerosol and SO$_2$ (green dot) under these conditions. Also shown is a line representing the range of possible solutions for an error-free measurement of X and Ka band absorptivity. For each possible SO$_2$ abundance, there is a corresponding cloud bulk density that can match the absorptivity measurement with equal probability, i.e. the set of possible solutions is infinite (but bounded). Although the SO$_2$ opacity model is linearized, we find that calculations of the probability distribution shown in Figure \ref{fig:degen} using the spectral line model for SO$_2$ exhibit negligible differences. It is therefore necessary to incorporate additional information, such as vertical structure assumptions, to arrive at a plausible simultaneous solution for H$_2$SO$_4$ aerosol and SO$_2$ retrievals.

\subsection{Model Atmospheres} \label{sec:mod_atm}
\begin{figure}[t]
\begin{center}
    \begin{minipage}[c]{0.48\textwidth}
    \centering
    \includegraphics[width=\textwidth]{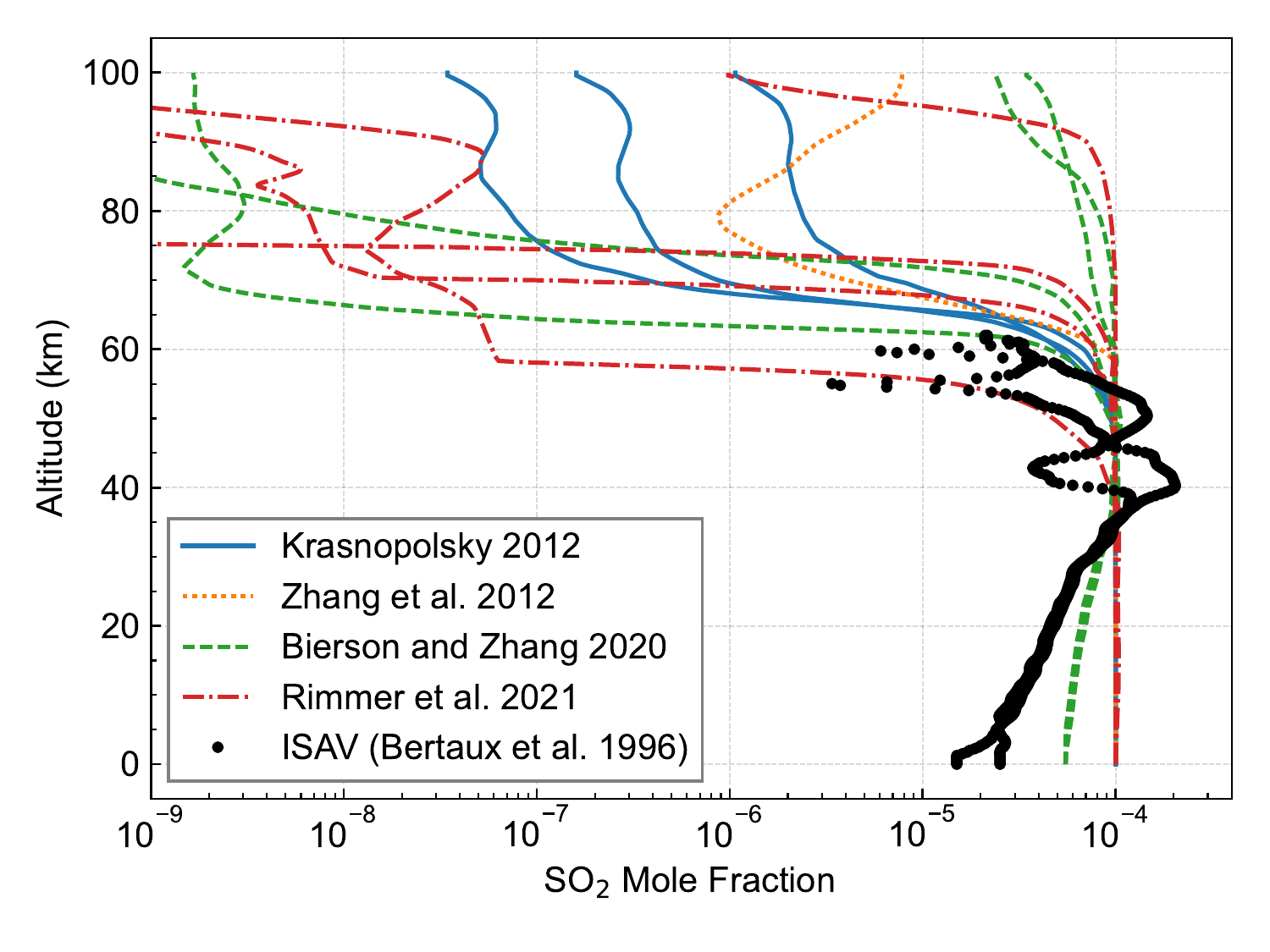}
    \end{minipage}
    \begin{minipage}[c]{0.48\textwidth}
    \centering
    \includegraphics[width=\textwidth]{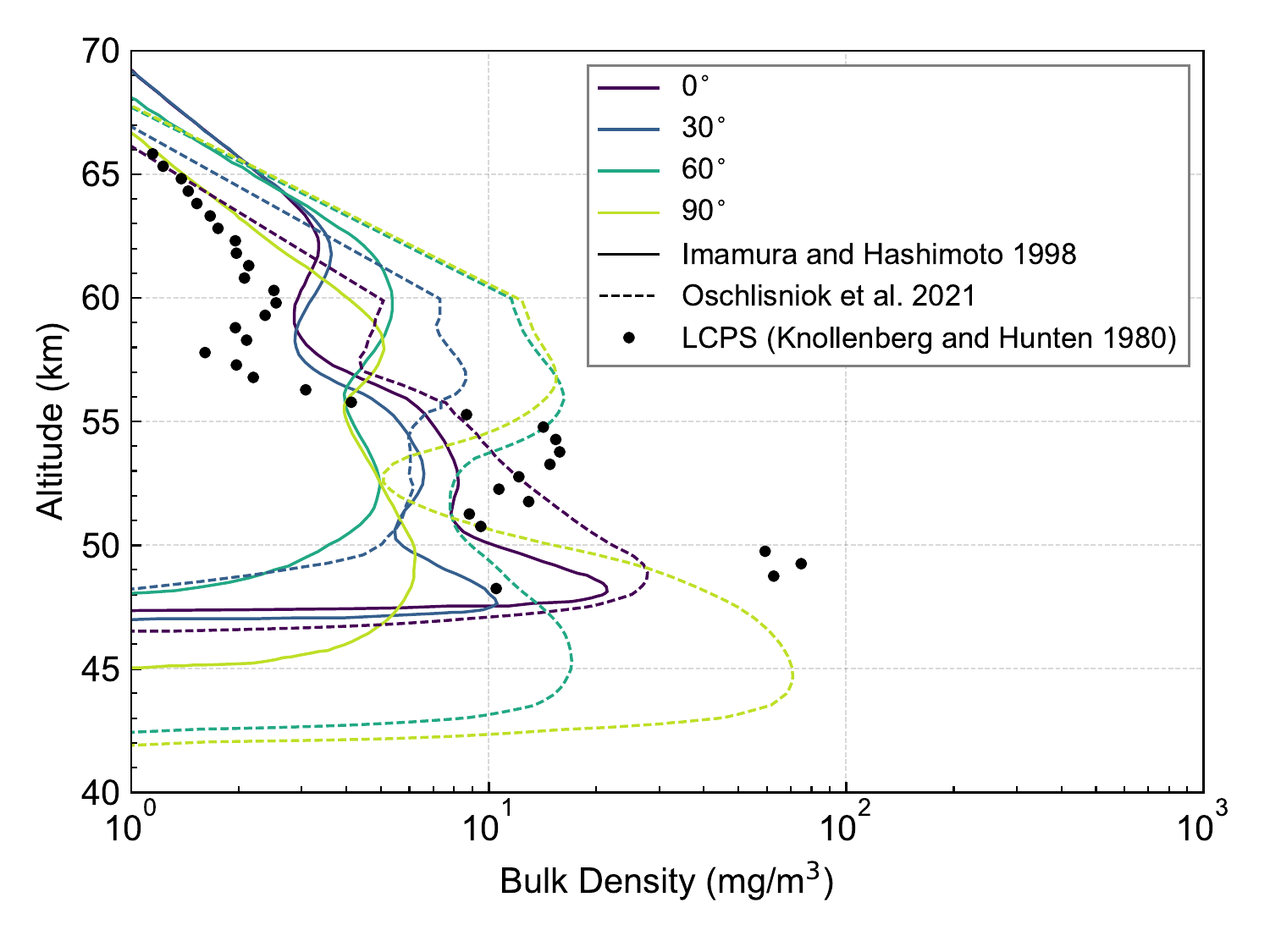}
    \end{minipage}
\caption{(Left) SO$_2$ mole fraction profiles from the chemical models of \citet{Krasnopolsky2012, Zhang2012, Bierson2020, Rimmer2021} and Vega ISAV spectrometer measurements \citep{Bertaux1996}. (Right) Cloud bulk density profiles from 2D transport models of \citet{Imamura1998, Oschlisniok2021} and Pioneer Venus LCPS measurements \citep{Knollenberg1980}}
\label{fig:so2_cloud}
\end{center}
\end{figure}

For our simulated retrievals, we consider different sets of model atmospheres defined in Table \ref{tab:datset}. Sets 1-4 include a range of different atmospheric profiles from sources which are not necessarily physically consistent. For atmospheric temperature and pressure profiles, we use latitude dependent model profiles derived from analysis of Venus Express and Akatsuki RO data \citep{Ando2020a}. Profiles of H$_2$SO$_4$ vapor derived from Venus Express radio occultations at several latitudes are used as provided by the VeRa team (Oschlisniok, personal communication). Since prior radio occultations and microwave/infrared imaging results have only suggested values for uniform sub-cloud abundance \citep{Oschlisniok2021, Jenkins2002, Arney2014}, our only sources of information on the vertical distribution of SO$_2$ are in situ measurements and chemical model predictions. Figure \ref{fig:so2_cloud} shows a collection of SO$_2$ profiles adjusted to a common base abundance of 100 ppm derived from contemporary chemical models \citep{Krasnopolsky2012, Zhang2012, Bierson2020, Rimmer2021} and from the in situ results of the Vega descent probe ISAV spectrometers \citep{Bertaux1996}. Specifically, we use the nominal profile of \citet{Krasnopolsky2012} representing conventional chemical model predictions, the cloud-layer inhibited transport model of \citet{Bierson2020}, and the cloud droplet depletion model of \citet{Rimmer2021}. The ISAV measurements deviate significantly from equilibrium chemical models, which indicate a uniform sub-cloud SO$_2$ abundance and limited gradients within the clouds themselves. Vertical structure information for the clouds is lacking in a similar sense and must also be considered in the context of modeling results and in situ data. Figure \ref{fig:so2_cloud} also shows a collection of cloud bulk density at varying latitudes from the 2D transport models of \citet{Imamura1998} and \citet{Oschlisniok2021}, as well as the Pioneer Venus LCPS measurements \citep{Knollenberg1980}. The model atmospheres in Sets 5-8 are based on the 2D transport model results of \citet{Oschlisniok2020,Oschlisniok2021} at 0, 40, and 80 degree latitude for temperature and H$_2$SO$_4$ abundances. Since SO$_2$ abundances are not solved by this model, the SO$_2$ profiles from Sets 1-4 are also used in Sets 5-8 .For each of the model atmospheres, vertical absorptivity profiles are determined at 8 and 32 GHz using the opacity models discussed in the previous section. Random covariant noise is added to these profiles using the statistical uncertainty matrices derived for each profile (e.g. Figure \ref{fig:abs_covar}, see Section \ref{sec:link_atten}) at a resolution of 0.5 km. 

\begin{table}[]
    \centering
    \begin{tabular}{cccc}
    Set & H$_2$SO$_4$ vapor & H$_2$SO$_4$ aerosol & SO$_2$  \\ \hline
    1 & VEX VeRa, 18$^\circ$ & \citet{Imamura1998}, 0$^\circ$ & \citet{Krasnopolsky2012} \\ 
    2 & VEX VeRa, 45$^\circ$ & \citet{Imamura1998}, 30$^\circ$ & \citet{Bierson2020} \\ 
    3 & VEX VeRa, 80$^\circ$ & \citet{Oschlisniok2021}, 60$^\circ$ & \citet{Rimmer2021} \\ 
    4 & VEX VeRa, 85$^\circ$ & \citet{Oschlisniok2021}, 90$^\circ$ & \citet{Bertaux1996}, ISAV-1 \\ \hline
    
    5 & \citet{Oschlisniok2021}, 0$^\circ$ & \citet{Oschlisniok2021}, 0$^\circ$ & \citet{Krasnopolsky2012} \\ 
    6 & \citet{Oschlisniok2021}, 40$^\circ$ & \citet{Oschlisniok2021}, 40$^\circ$ & \citet{Bierson2020} \\ 
    7 & \citet{Oschlisniok2021}, 80$^\circ$ & \citet{Oschlisniok2021}, 80$^\circ$ & \citet{Rimmer2021} \\ 
    8 & \citet{Oschlisniok2021}, 40$^\circ$ & \citet{Oschlisniok2021}, 40$^\circ$ & \citet{Bertaux1996}, ISAV-1 \\ \hline
    \end{tabular}
    \caption{Latitude-dependent ground-truth atmospheric profiles for simulated retrievals.}
    \label{tab:datset}
\end{table}

\subsection{Profile Retrieval Approaches}
Least-squares minimization of Equation \ref{eqn:lgprb} results in a maximum likelihood estimation of the abundance profiles. Since we are now considering profile retrievals, the definitions of the $\mathbf{x}$, $\mathbf{y}$, $\mathbf{K}$, and $\mathbf{S_y}$ matrices introduced in Section \ref{sec:ill-pose} are now expanded to include the full profiles. For a uniform altitude grid of length $n$, the length of $\mathbf{x}$ becomes $3n$, and the length of $\mathbf{y}$ becomes $2n$. To minimize Equation \ref{eqn:lgprb}, we use the \verb|scipy.optimize| package implementation of Powell's method \citep{2020SciPy-NMeth}. An estimate for the covariance of the retrieved profiles $\mathbf{S_x}$ without regularization conditioning can be found from the pseudoinverse of the transformed measurement error covariance. 

\begin{equation} \label{eq:pseudo}
\mathbf{\hat{S}_x} = (\mathbf{K^TS^{-1}_y} \mathbf{K})^+  
\end{equation}

\begin{figure}
    \centering
    \includegraphics[width=\textwidth]{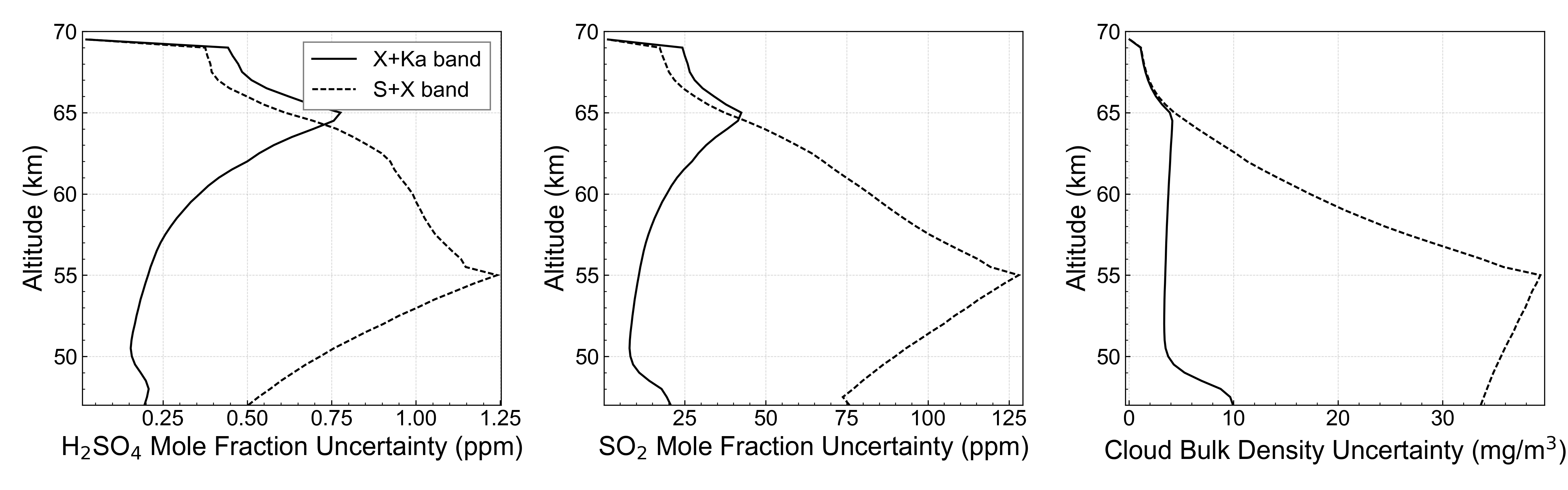}
    \caption{Minimum uncertainty estimates (see text) for X/Ka band RO retrievals of H$_2$SO$_4$ and SO$_2$ abundances at 0.5 km resolution compared to S/X band retrievals assuming other absorber abundances are known exactly. Inflection points occur at altitudes where the statistical uncertainty exceeds the abundance necessary to match the Set 2 model atmosphere absorptivities.}
    \label{fig:ex_mse_er}
\end{figure}

This inversion, however, is highly ill-conditioned for joint estimates of neutral species abundances. Figure \ref{fig:ex_mse_er} shows the uncertainties associated with this pseudoinverse for retrievals of H$_2$SO$_4$ and SO$_2$ abundances under the limiting assumption that, for each of the absorbing species, the abundances of the other absorbers are known exactly. This sets a lower limit on the achievable uncertainties at 0.15 ppm for H$_2$SO$_4$ vapor, 8 ppm for SO$_2$, and 3.5 mg/m$^3$ for H$_2$SO$_4$ aerosol. These uncertainty estimates for X and Ka band retrievals determined in this way are significantly lower than those for S and X band retrievals; this illustrates why previous attempts at joint retrieval of H$_2$SO$_4$ vapor and SO$_2$ from S and X band RO measurements directly (instead of using the saturation depletion assumption) have yielded unlikely results \citep{Jenkins1994}.
 
Since the simultaneous retrieval problem is under-constrained and ill-posed, minimization of Equation \ref{eqn:lgprb} is strongly dependent on the starting guesses (or seed profiles) provided to the solver and the regularization (i.e.  conditioning, with terms used interchangeably) strategy. We propose a multi-step approach to seed and regularize the problem, as discussed in the following sections. The inputs for the retrieval are the temperature and pressure profiles derived from the RO Doppler shift measurements, X and Ka band absorptivities, and estimates of the absorptivity per-band covariance. The outputs are abundance profiles for H$_2$SO$_4$ and SO$_2$ species at 0.5 km resolution from the Ka band attenuation limit (45-50 km) to 70 km and estimations of the retrieval uncertainties.

\subsubsection{Atmospheric transport model}
\begin{figure}
    \centering
    \includegraphics[width=0.8\textwidth]{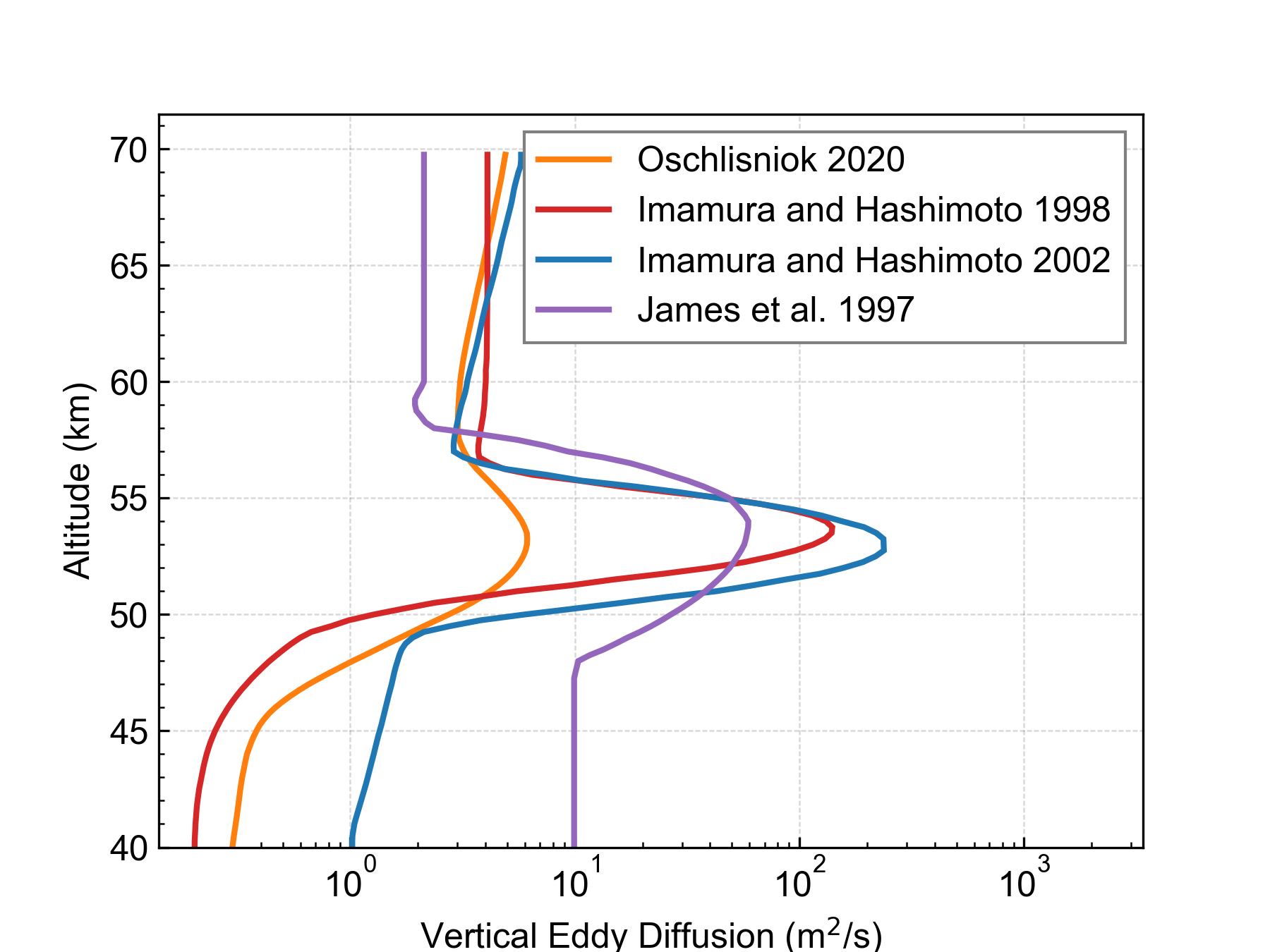}
    \caption{Eddy diffusion profiles used in the 1D atmospheric transport model }
    \label{fig:eddy_diffuse}
\end{figure}

\begin{figure}
    \centering
    \includegraphics[width=0.8\textwidth]{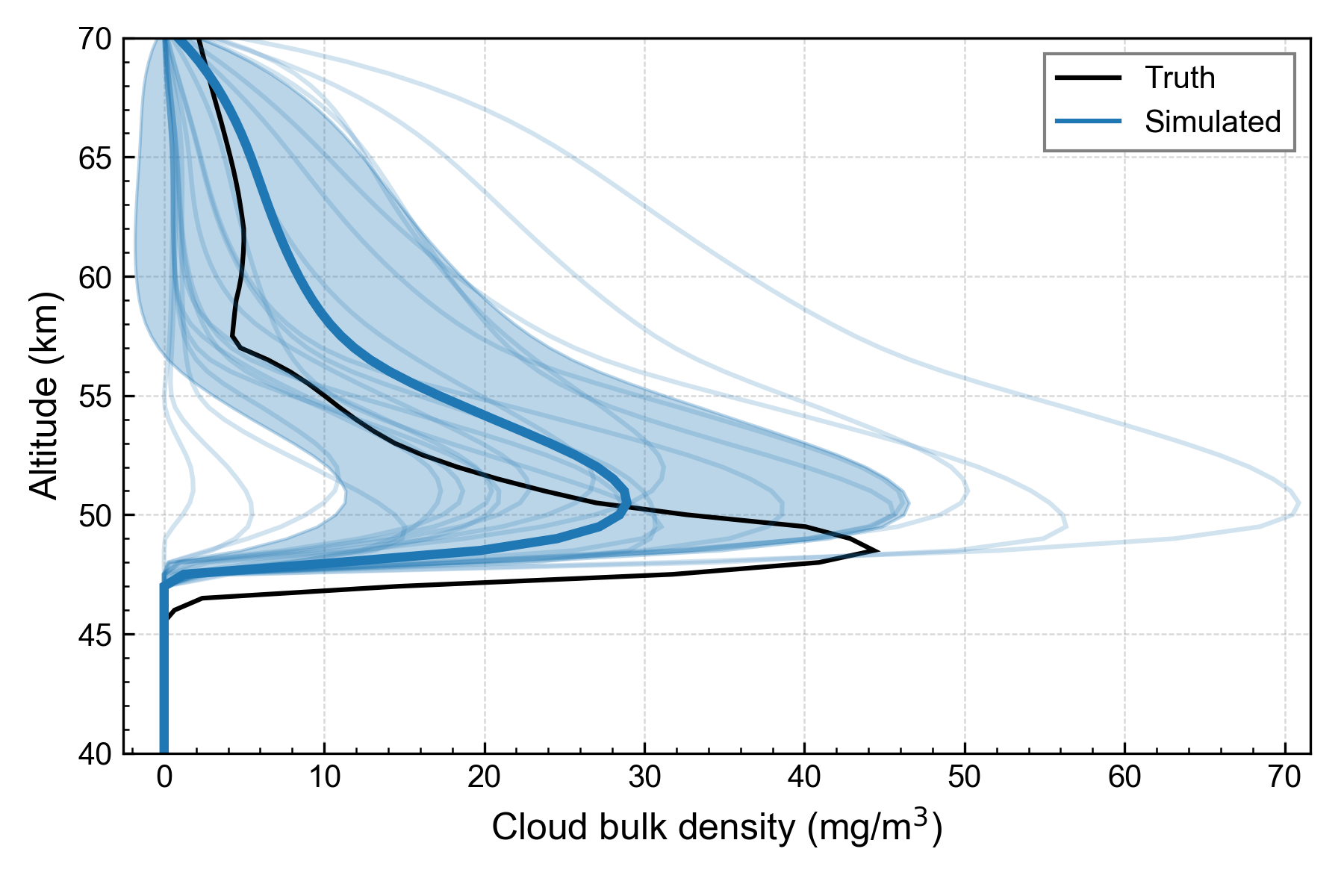}
    \caption{Results of cloud aerosol mass simulations for the Set 5 model atmosphere. The ensemble of simulations is used to derive a mean profile and associated uncertainties, as shown compared to the true profile.}
    \label{fig:cloud_model}
\end{figure}

 First, an initial estimate of the H$_2$SO$_4$ vapor abundance is determined from the X band profile to its attenuation limit via direct inversion of the opacity equation (i.e. assuming all opacity is due to H$_2$SO$_4$ vapor). The initial determination of the H$_2$SO$_4$ vapor profile is equivalent in precision to profiles derived from prior single-band radio occultation measurements and will somewhat overestimate the vapor abundance. This initial H$_2$SO$_4$ vapor estimate and the retrieved temperature are used as inputs to a 1D transport model of the H$_2$SO$_4$ aerosol system to develop estimates for cloud bulk density. We use a simplified 1D advection-diffusion transport model based on the previously published 1D cloud microphysics \citep{James1997, Imamura2001, McGouldrick2007} and 2D transport \citep{Imamura1998, Oschlisniok2021} models of Venus' H$_2$SO$_4$ aerosol system. The active physical processes in this model are eddy diffusion, sedimentation of cloud aerosols, mean vertical winds, and cloud condensation/vaporization. We adopt the nominal aerosol sedimentation velocity profiles of \citet{Oschlisniok2021}, and we use H$_2$SO$_4$ vapor pressure laws suggested by \citet{Krasnopolsky2015} assuming a constant cloud weight percent profile \citep{Hashimoto2001}.
Our model ignores the impact of cloud microphysics; it is assumed that the distribution of H$_2$SO$_4$ between vapor and liquid phases is governed solely by the saturation vapor pressure. Microphysical processes are omitted because other models have shown that H$_2$SO$_4$ vapor abundance in Venus' atmosphere follows the saturation vapor pressure curve for nominal abundances of cloud condensation nuclei \citep{James1997, McGouldrick2007}. Additionally, X and Ka band link attenuation by the clouds is only sensitive to the total cloud mass since the nominal diameters of Venus cloud aerosols are well within one wavelength (i.e. they are negligible scatterers). We also do not consider the effect of H$_2$SO$_4$ photochemical production, or H$_2$SO$_4$ thermal dissociation. We solve the advection-diffusion equations using a Semi-Lagrangian Crank-Nicholson finite difference scheme \citep{Spiegelman2006} with Neumann boundary conditions ($\frac{\partial n}{\partial x} = 0$). The intial guess H$_2$SO$_4$ vapor profile is held constant throughout the simulation. The model is run until convergence (over several Venus years), which yields the initial estimate for the cloud profile.  Uncertainties in the cloud bulk density are estimated by varying the eddy diffusion coefficients and mean vertical winds for several model runs. We use a range of eddy diffusion coefficients implemented in previously published models, as shown in Figure \ref{fig:eddy_diffuse}. A resulting simulation is shown compared to the Set 5 model atmosphere in Figure \ref{fig:cloud_model}.  

\subsubsection{MCMC Shape Model Fitting}

After determining the mean cloud profile estimate from the transport model, the initial guess for H$_2$SO$_4$ vapor is refined by fitting to both X and Ka band absorptivity profiles, which improves the estimate over the final retrieval altitude range. Next, a Markov Chain Monte Carlo (MCMC) approach is used to perform a parametric fit the X and Ka band absorptivity using the output of the cloud model and a shape model for SO$_2$. Three parameters are used to define the SO$_2$ vertical abundance profile: the base abundance $q_0$, the depletion altitude $h_0$, and the depletion scale height $s$. The cloud model output is additionally scaled, leading to a total of 4 free parameters.

\begin{equation}
q_{SO_2}(h) = 
\begin{cases}
    q_0 & h \leq h_0  \\ 
    q_0e^{-(h - h_0)/s} & h > h_0
\end{cases}
\end{equation}

\begin{figure}
    \centering
    \includegraphics[width=0.8\textwidth]{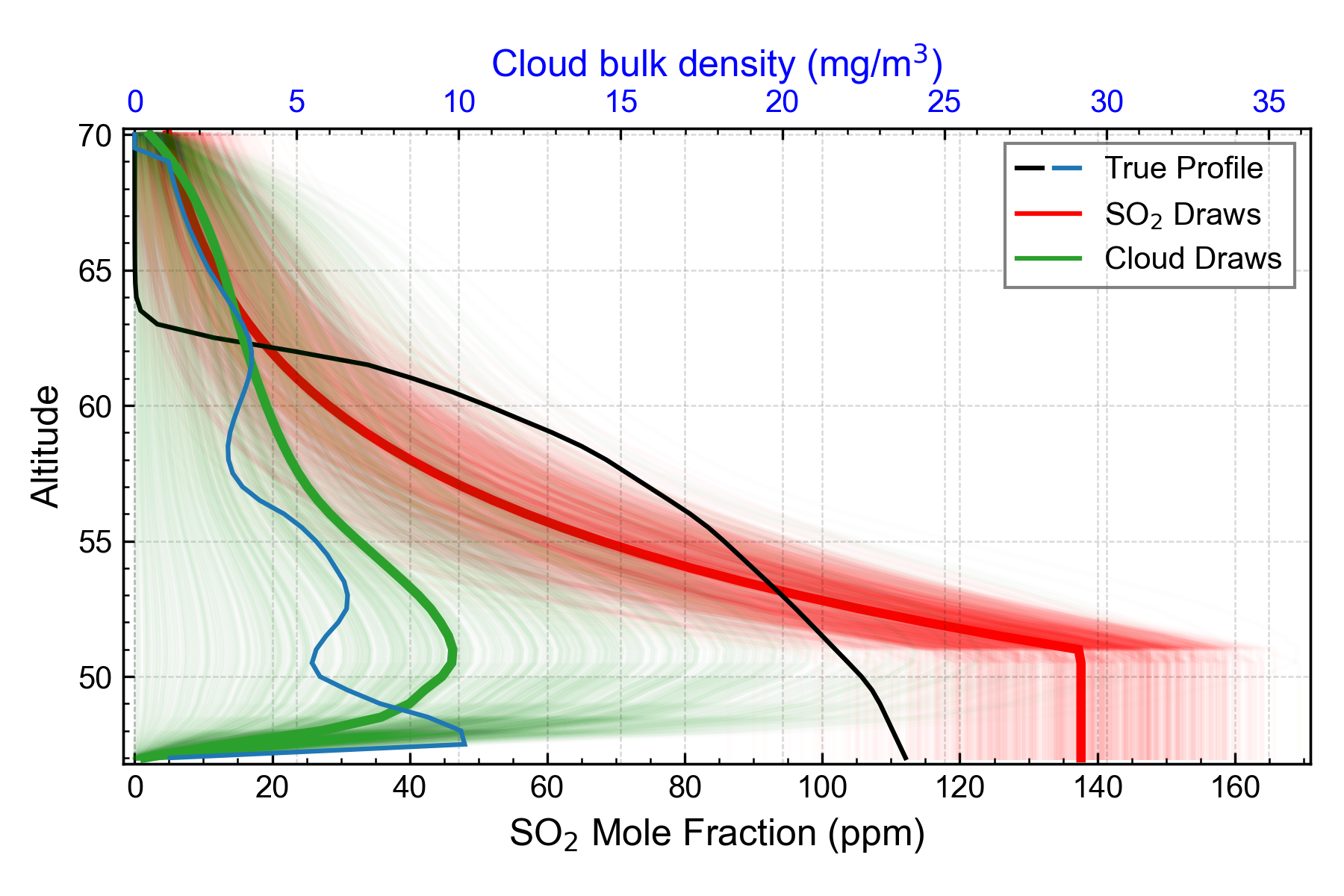}
    \caption{Draws from the posterior distribution and the median results for H$_2$SO$_4$ aerosol and SO$_2$ profiles for the Set 2 model atmosphere.}
    \label{fig:mcmc_draws}
\end{figure}

The MCMC approach provides an initial estimate for the profile shape by estimating the likelihood distribution $P(x|y)$ for each model parameter (see \citet{Foreman-Mackey2013} for a discussion of MCMC estimation). This fitting method is useful because in addition to providing a seed profile for $\mathbf{x}$, the collection of sampled profiles in the converged distribution also provides a preliminary estimate of variance in the retrieved quantities. Each MCMC fit executes 10000 iterations plus 500 burn-in steps for sets of 100 walkers to arrive at the final likelihood distribution. An example of the fit results using this procedure is shown in Figure \ref{fig:mcmc_draws} for the Set 2 model atmosphere. 

\subsubsection{Conditioned retrieval}
 Both prior steps provide initial estimates for the H$_2$SO$_4$ and SO$_2$ abundances based on assumptions as to the general shape of these profiles. These profile estimates are then used as seed profiles $x\mathbf{x_a}$ for a regularized minimization of Equation \ref{eqn:lgprb}. We use the following form for the regularization term $\mathcal{R}$ which is added to Equation \ref{eqn:lgprb}. 

\begin{equation} \label{eq:reg_eq}
  \mathcal{R} = b_1 (\mathbf{x} - \mathbf{x_a})^T \mathbf{\hat{S}_a^{-1}} (\mathbf{x} - \mathbf{x_a}) +  b_2 \mathbf{x}^T \mathbf{\Gamma \Gamma}^T \mathbf{x}
\end{equation}

While the regularization applies only to the SO$_2$ and H$_2$SO$_4$ aerosol profiles, we retain the matrix notation for convenience. The first term of Equation \ref{eq:reg_eq} penalizes deviation from the seed profiles. We note that while this resembles the form of a retrieval incorporating a priori information, the $\mathbf{x_a}$ seed profiles are not true a priori, since they were determined from fits to the data under strict assumptions \citep{Rodgers2000}. In this expression, the diagonal $\mathbf{\hat{S}_a^{-1}}$ matrix is the inverse of the variance from the MCMC step. The $\mathbf{\Gamma}$ matrix in the second term represents the application of high pass filter to the retrieved SO$_2$ and H$_2$SO$_4$ aerosol abundance profiles. Specifically, $\mathbf{\Gamma}$ represents a finite impulse response (FIR) filter matrix. The constant terms $b$ are used to weight the regularization and are determined empirically. The addition of the regularization terms to Equation \ref{eqn:lgprb} also modifies the corresponding estimate of uncertainty originally stated in Equation \ref{eq:pseudo}. These regularization terms condition the matrix inverse, and the resulting inversion provides useful estimates of retrieval uncertainties.

\begin{equation} \label{eq:pseudo2}
\mathbf{\hat{S}_x} = (\mathbf{K^TS^{-1}_y} \mathbf{K} + b_1 \mathbf{\hat{S}_a^{-1}} + b_2 \mathbf{\Gamma \Gamma}^T)^+  
\end{equation}

\subsection{Simulation Results} 
\begin{figure}[htbp]
\begin{center}
\includegraphics[width=\textwidth]{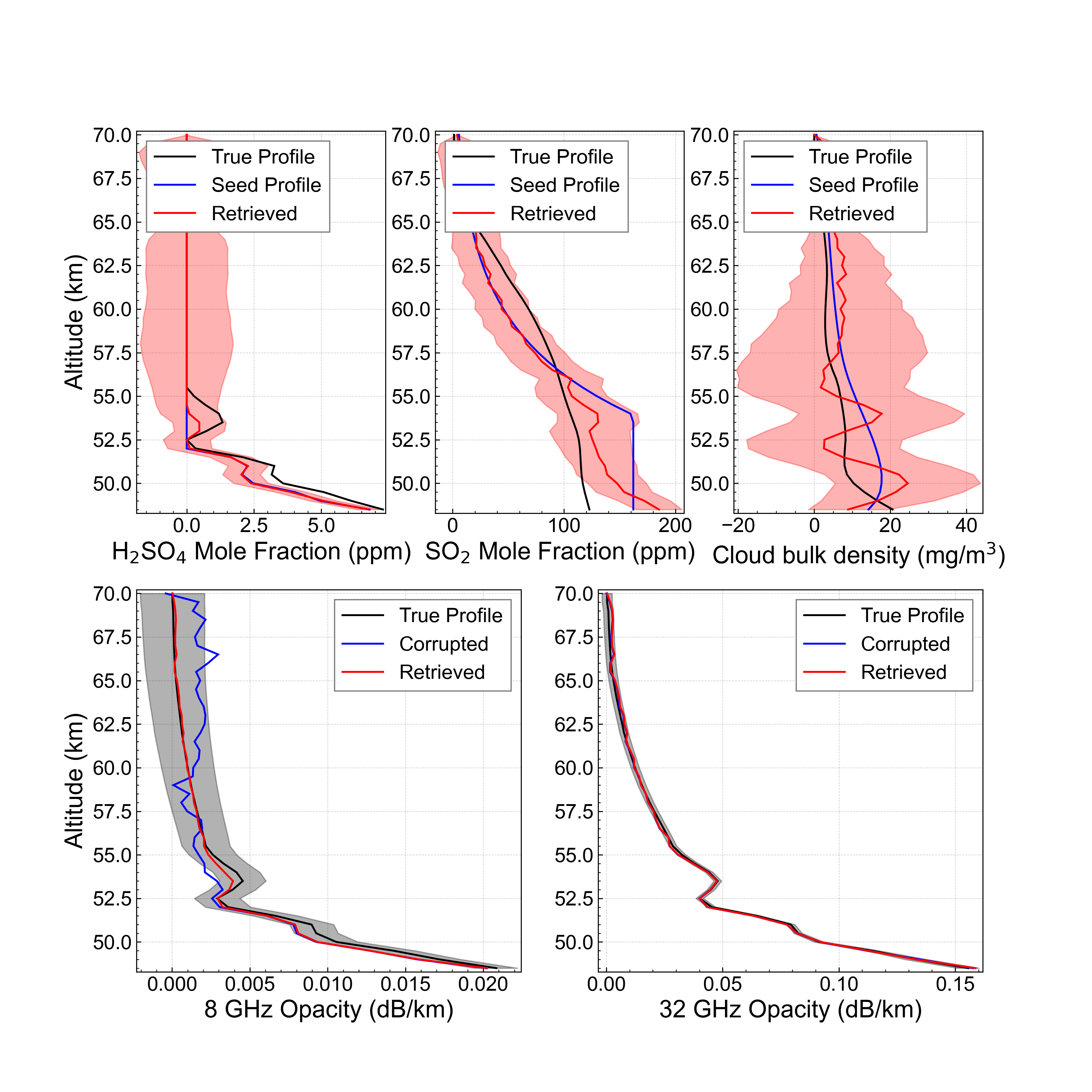}
\caption{Abundances of H$_2$SO$_4$ and SO$_2$ (top) retrieved from simulated dual X/Ka band RO absorptivities (bottom) for the Set 1 model atmosphere, with uncertainties determined using Equation \ref{eq:pseudo2}. Seed profiles were provided to the final optimization stage from the outputs of the atmospheric transport and MCMC steps (see text).}
\label{fig:inp_out}
\end{center}
\end{figure}

\begin{figure}[t]
\begin{center}
\begin{minipage}[c]{0.48\textwidth}
\centering
\textbf{Set 1} \par
\includegraphics[width=\textwidth]{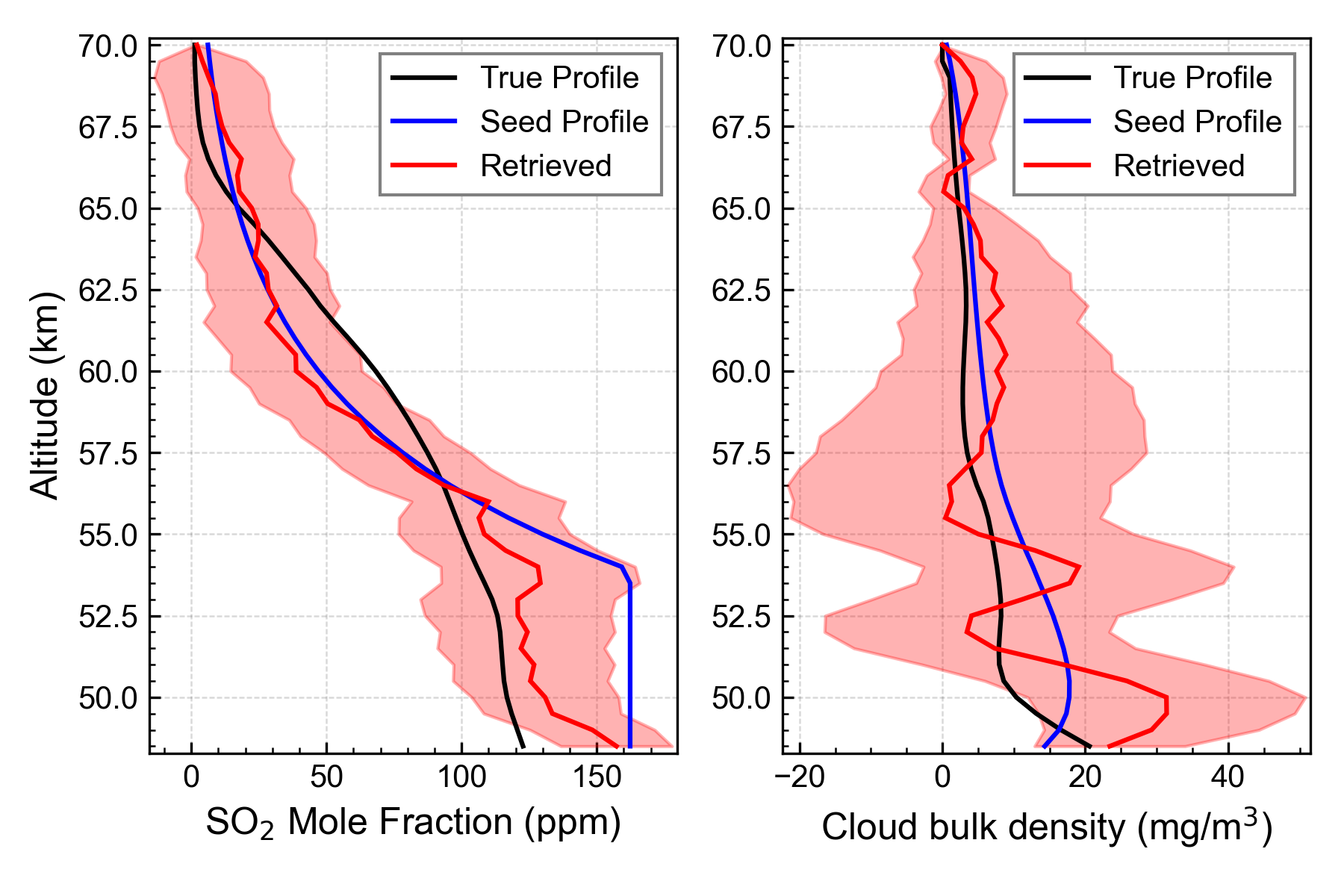}
\end{minipage} 
\hfill
\begin{minipage}[c]{0.48\textwidth}
\centering
\textbf{Set 2} \par
\includegraphics[width=\textwidth]{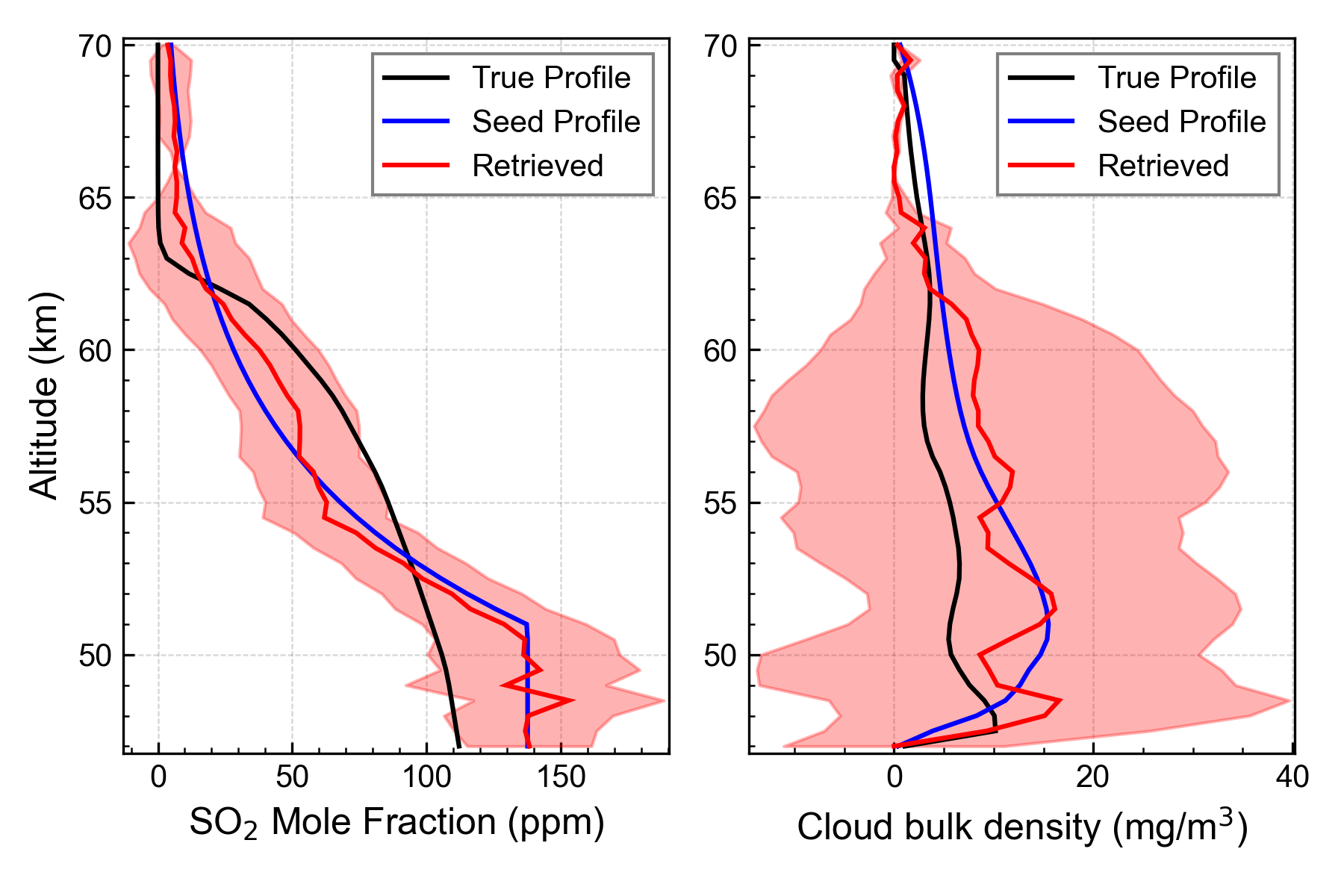}
\end{minipage}
\begin{minipage}[c]{0.48\textwidth}
\centering
\textbf{Set 3} \par
\includegraphics[width=\textwidth]{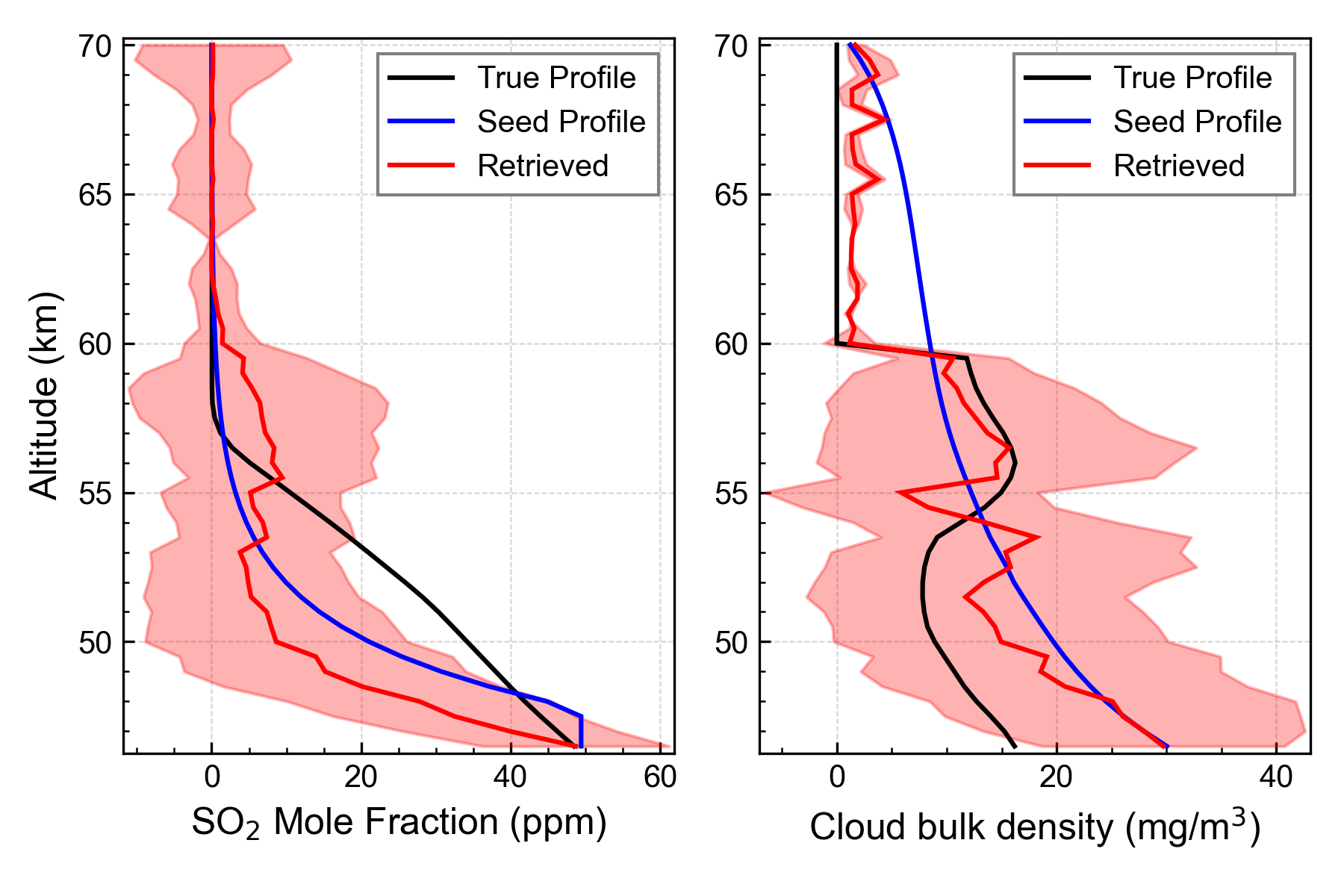}
\end{minipage} 
\hfill
\begin{minipage}[c]{0.48\textwidth}
\centering
\textbf{Set 4} \par
\includegraphics[width=\textwidth]{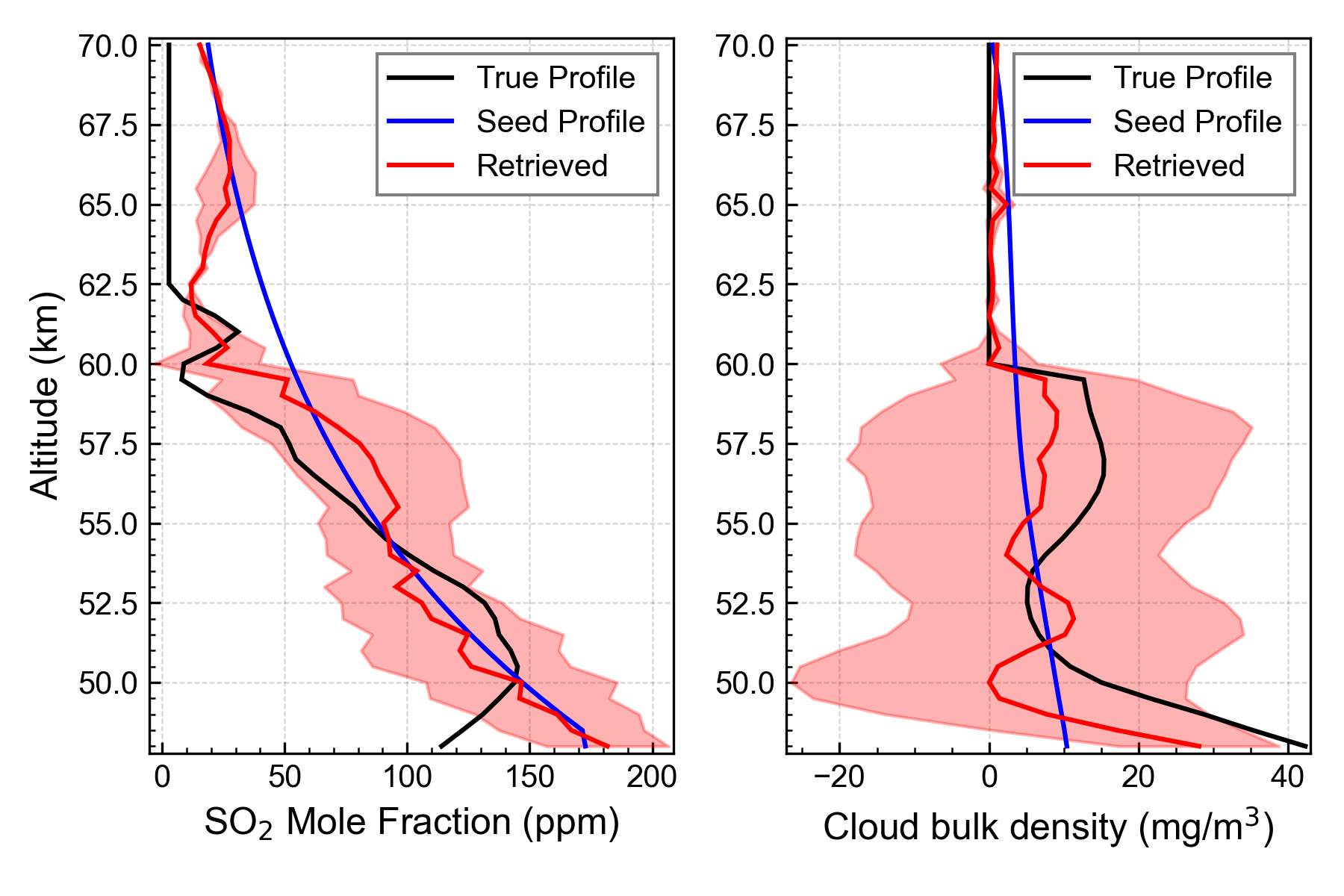}
\end{minipage} 
\begin{minipage}[c]{0.48\textwidth}
\centering
\textbf{Set 5} \par
\includegraphics[width=\textwidth]{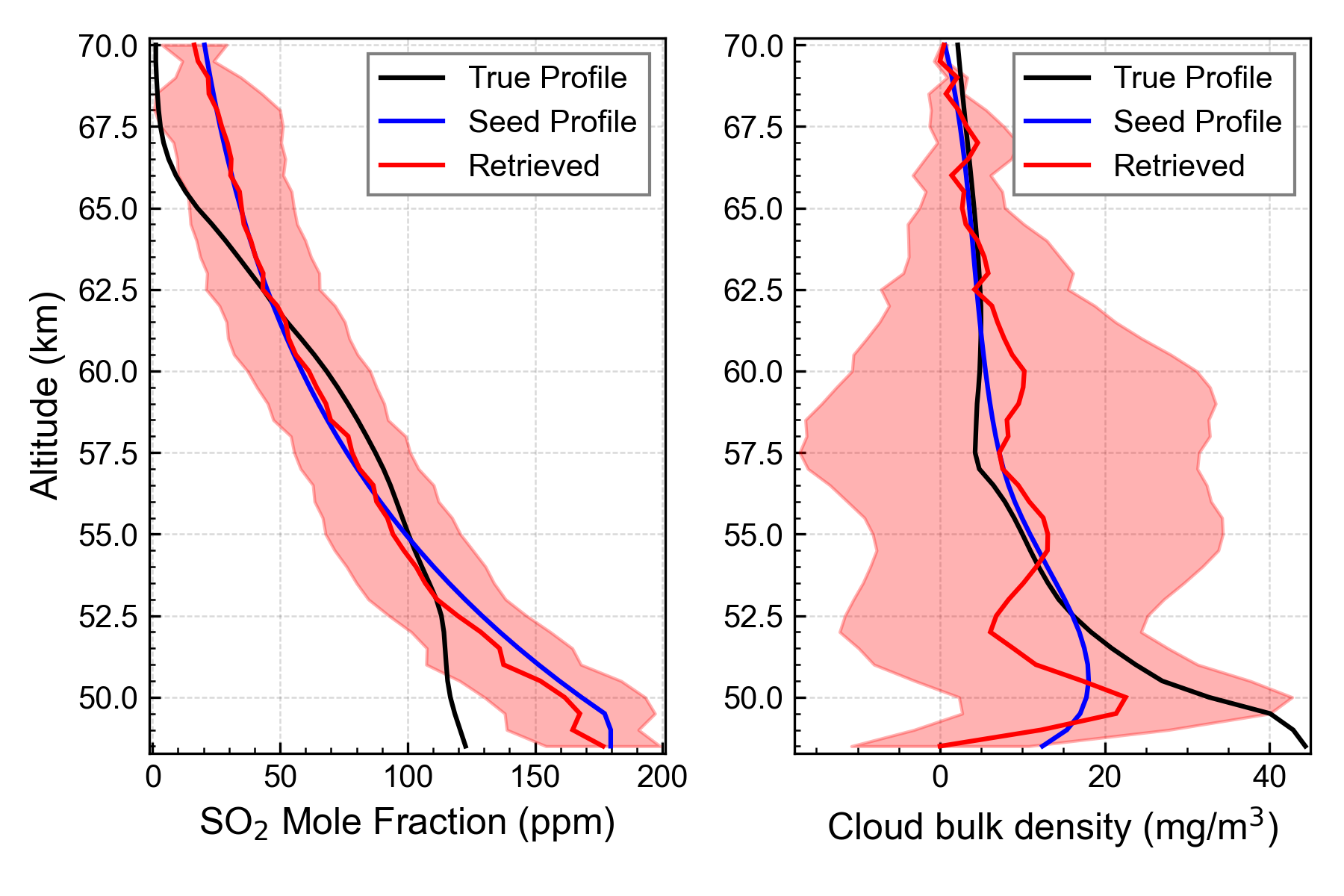}
\end{minipage} 
\hfill
\begin{minipage}[c]{0.48\textwidth}
\centering
\textbf{Set 6} \par
\includegraphics[width=\textwidth]{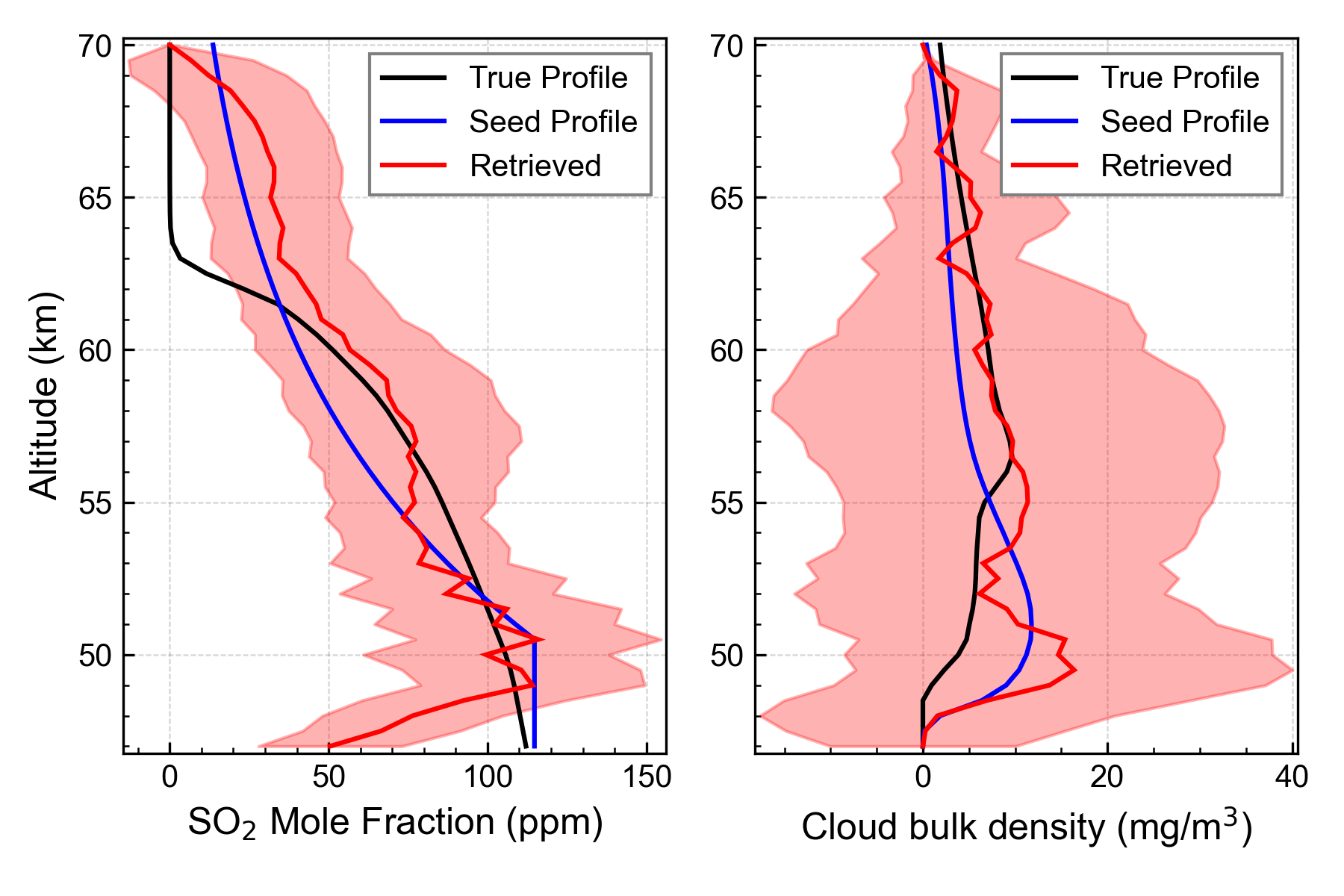}
\end{minipage}
\begin{minipage}[c]{0.48\textwidth}
\centering
\textbf{Set 7} \par
\includegraphics[width=\textwidth]{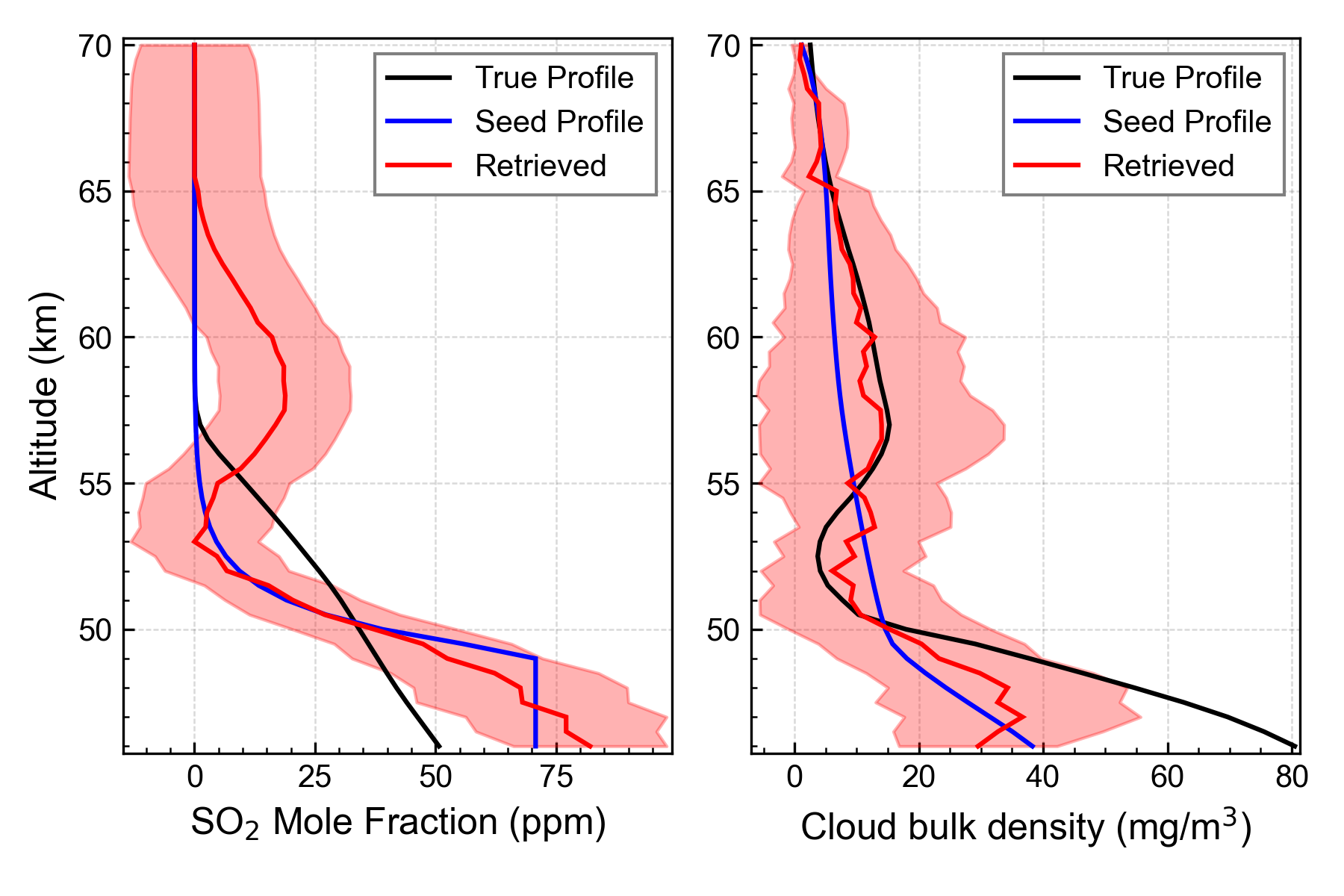}
\end{minipage} 
\hfill
\begin{minipage}[c]{0.48\textwidth}
\centering
\textbf{Set 8} \par
\includegraphics[width=\textwidth]{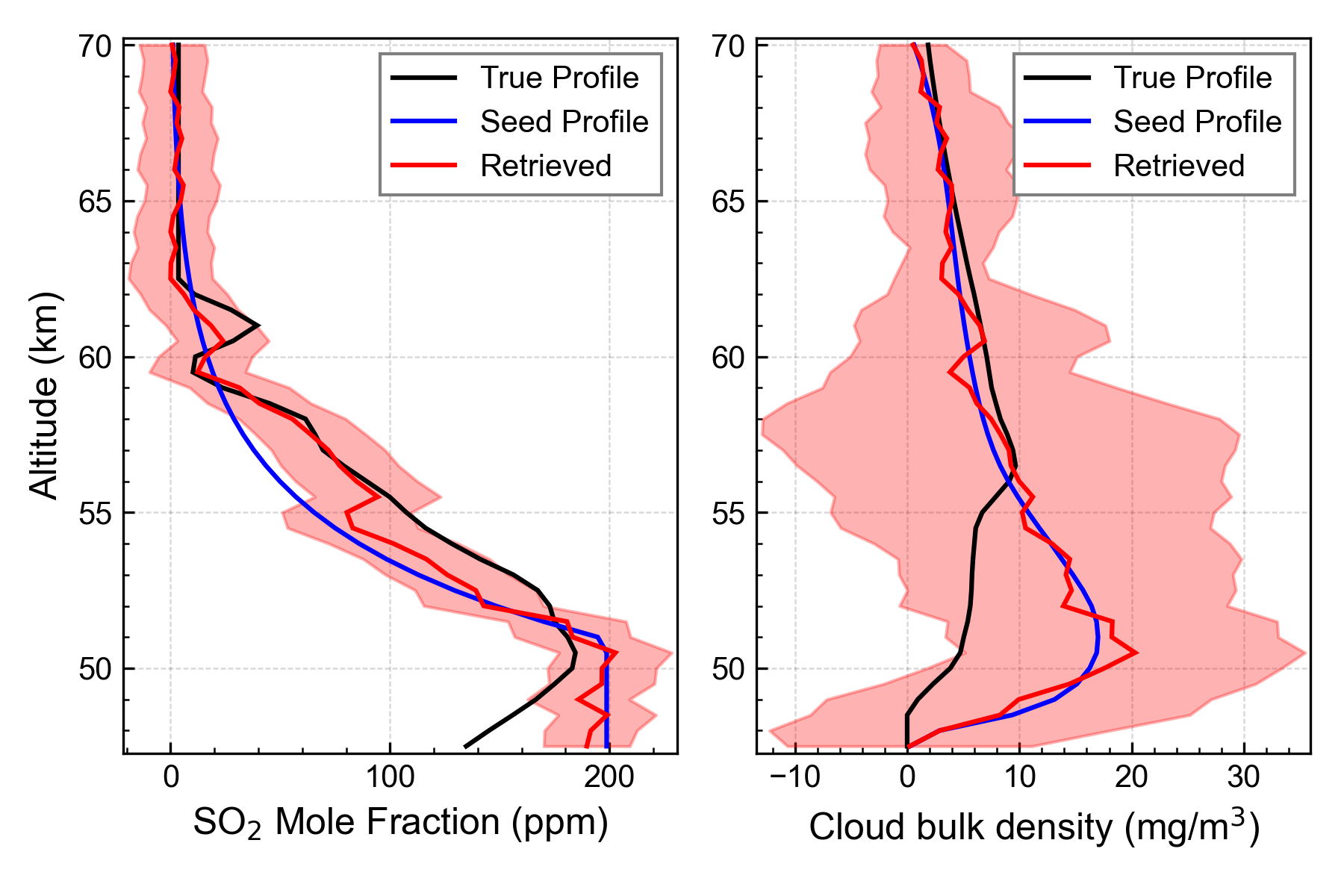}
\end{minipage} 
\caption{Retrievals of Set 1-8 model atmosphere abundances of SO$_2$ and H$_2$SO$_4$ aerosol with uncertainties.}
\label{fig:s18}
\end{center}
\end{figure}

To test the efficacy of this approach for retrieving SO$_2$ and H$_2$SO$_4$ aerosol abundances from dual X/Ka band RO measurements, we conducted simulated retrievals for the atmosphere models enumerated in Table \ref{tab:datset}. Figure \ref{fig:inp_out} illustrates the simulated retrieval inputs and outputs for the Set 1 model atmosphere. The vertical profiles of atmospheric neutrals shown in black in the top row were used to compute X and Ka band absorptivities, and noise is added to these profiles by adding samples from a multivariate normal distribution (the absorptivities at each altitude point represent random variables in the ensemble) with statistics specified by the corresponding simulation covariance matrix (e.g. Figure \ref{fig:abs_covar}). The corrupted absorptivities, shown in blue on the bottom row, are then used to retrieve the atmospheric profiles. The transport model provides initial estimates for cloud bulk density conditions over a range of different  advection and diffusion conditions. The mean cloud profile scale and parameters for the SO$_2$ shape model are then adjusted using the MCMC fitting procedure, and the outputs are used as seed profiles for the final retrieval at full resolution. The diagonal variance of the MCMC samples is used as the $\mathbf{\hat{S}_a}$ matrix since the full covariance output can be poorly conditioned and difficult to invert. Rather than using the output cloud profile from the MCMC step, the seed cloud profile was set by taking the mean of the scaled cloud profile and the original mean model output, which generally improved the resulting fits. We experimented with using an eigendecomposition of the resulting cloud model profiles to increase the number of free parameters for the cloud model in the MCMC fit but found that no significant improvement was observed over simply scaling the mean profile (which is similar in shape to the principal eigenvector). For the $b_1$ constant, we found a value of $0.005$ as a good empirical weight. Weights of this order of magnitude somewhat deprioritize agreement with the seed profile in the final result while being superior to a zero weight. The b$_2$ weights were set separately for the SO$_2$ and cloud profiles as the inverse of the maximum value of their seed profiles. The FIR filter highpass cutoff wavenumbers were determined empirically as 0.15 km$^{-1}$ and 0.25 km$^{-1}$ for the SO$_2$ and cloud profiles, respectively. Figure \ref{fig:s18} compares the retrieved abundances of SO$_2$ and H$_2$SO$_4$ aerosols with the true profile and seed profiles for the Set 1-8 model atmospheres. For the final retrieved profiles, the average profile mean (and maximum) errors below 55 km are 0.4 (0.7) ppm for H$_2$SO$_4$ vapor, 20 (47) ppm for SO$_2$, and 9 (24) mg/m$^3$ for H$_2$SO$_4$ aerosol. The corresponding average uncertainties estimated using Equation \ref{eq:pseudo2} are 0.7 ppm for H$_2$SO$_4$ vapor, 25 ppm for SO$_2$, and 18 mg/m$^3$ for H$_2$SO$_4$ aerosol. 

\begin{figure}
    \centering
    \includegraphics[width=\textwidth]{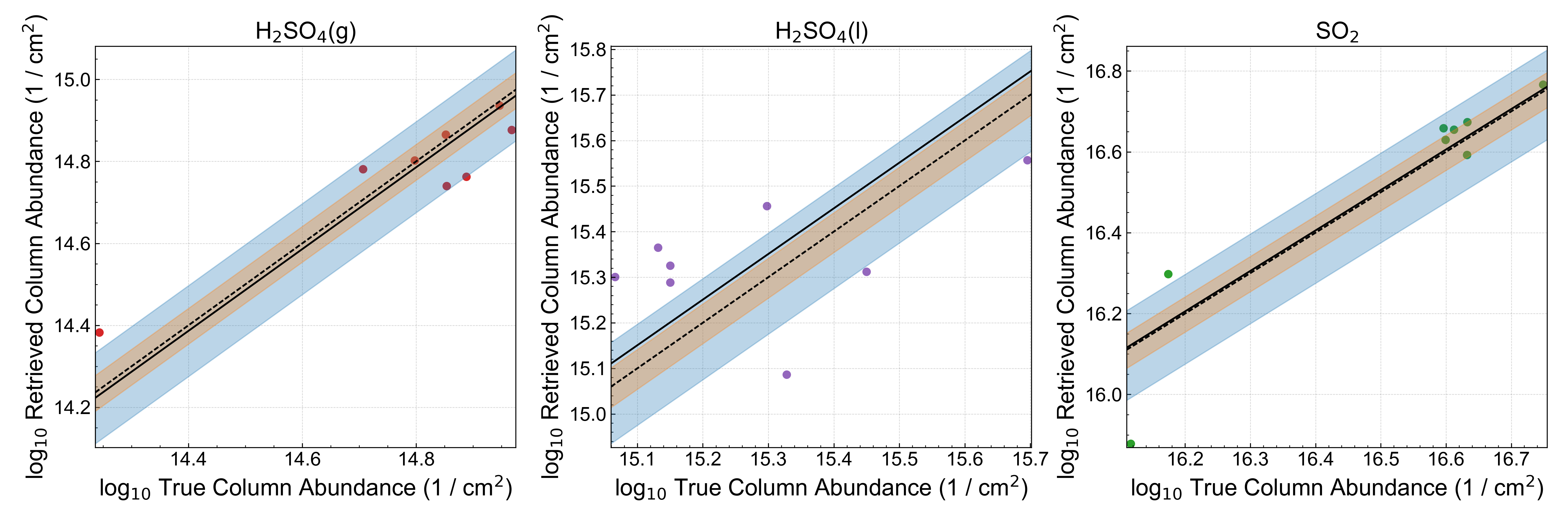}
    \caption{Comparison of retrieved and true column abundances of H$_2$SO$_4$ and SO$_2$. A linear model (black line) is fit to the data assuming a zero intercept and compared to a one-to-one relationship (dashed black line). Regions of 10\% (shaded orange) and 25\% (shaded blue) are also shown.}
    \label{fig:columns}
\end{figure}

In addition to retrieving profiles of SO$_2$ and H$_2$SO$_4$ aerosols, we also compared the retrieved column abundance of each species with its true value in Figure \ref{fig:columns}. The relationship between the retrieved and true column-integrated quantities can suggest potential biases in the retrieval, although no conclusions are drawn here due to the limited number of simulated retrievals. The dashed line indicates the region where the retrieval matches the true value ($y=x$), and a solid line indicates a best fit slope to the data assuming a zero intercept. Regions of 10\% and 25\% difference from the linear relationship are also shown. While a slight (~10\%) positive bias is apparent in the H$_2$SO$_4$ aerosol result, minimal bias is observed for SO$_2$ and H$_2$SO$_4$ vapor.

\section{Discussion} \label{sec:discussion}

In Section \ref{sec:link_atten}, we gave an overview of methods for computing uncertainties in radio occultation profiles. We used simple radio occultation simulations and radio link characteristics expected for EnVision measurements to compute the full absorptivity covariance matrices. These absorptivity uncertainties were used for all simulated retrievals presented here, and we consider them to be sufficiently characteristic of uncertainties that will likely be encountered during actual occultations. Of course, variations in transmitter signal strength, knowledge of antenna gain/pointing, occultation geometry, and uncertainty in spacecraft trajectory will all impact the resulting uncertainties; we refer the reader to \citet{Jenkins1994} and \citet{Oschlisniok2012,Oschlisniok2021} for a sense of the variability of these uncertainties at X band. Future work should incorporate realistic occultation geometries for retrieval simulations. We have also derived uncertainties in the opacity models for gases and aerosols in Venus' atmosphere based on laboratory measurements. We used the raw data tables supplied with the papers describing these laboratory measurements, which included estimates of random and systematic uncertainties. Of these neutral species, the measurements of H$_2$SO$_4$ vapor have the greatest uncertainty due to the considerable difficulty in making accurate laboratory measurements under simulated Venus conditions (see \citet{Kolodner1998a, Akins2019, Akins2020a} for details of those experiments). While we exclude these uncertainties from our simulated retrievals out of a desire to isolate the uncertainties most closely associated with the retrieval approach, they will need to be taken into account in the analysis of future dual X/Ka band RO measurements. 

In Section \ref{sec:retmet}, we have illustrated the prospects for the simultaneous retrieval of H$_2$SO$_4$ vapor, H$_2$SO$_4$ aerosol, and SO$_2$ abundance profiles in Venus' atmosphere using X and Ka band absorptivity profiles. As with earlier RO experiments, the retrieval of H$_2$SO$_4$ vapor from such measurements is on firm ground. Simultaneous retrieval at X and Ka band should achieve accuracies similar to prior dual S/X band occultations \citep{Jenkins1994}. Retrievals of SO$_2$ and H$_2$SO$_4$ aerosol are more uncertain and require careful regularization. Previous attempts to retrieve SO$_2$ abundances from Venus RO measurements have taken two approaches. The first is to retrieve both abundances simultaneously from S and X band measurements. \citet{Jenkins1994} used Magellan S and X band absorptivity profiles for orbit 3212 to solve for H$_2$SO$_4$ vapor and SO$_2$ simultaneously, finding SO$_2$ uncertainties ranging from 50 ppm in H$_2$SO$_4$ vapor-free regions to 200 ppm in regions where vapor was present. The perturbative approach used by \citet{Jenkins1994} to determine SO$_2$ uncertainties is similar to the approach we employed to determine our Figure \ref{fig:ex_mse_er} using the Set 2 model atmosphere, and the uncertainty estimates are consistent when converted to equivalent vertical resolutions (1 km for \citet{Jenkins1994} vs 500 m resolution in Figure \ref{fig:ex_mse_er})\footnote{We note that the SO$_2$ retrieval in Figure 10 of \citet{Jenkins1994} is biased high due to the use of an older H$_2$SO$_4$ vapor opacity model (see \citet{Kolodner1998a})}. The second approach is to assume that the H$_2$SO$_4$ vapor profile above the cloud base agrees well with the saturation vapor pressure, i.e. H$_2$SO$_4$ vapor at higher altitudes is depleted efficiently via condensation. We note that a similar justification is used to ignore the effects of cloud microphysics and condensation nuclei availability in our atmospheric transport model. Following this assumption, \citet{Oschlisniok2021} determined an SO$_2$ abundance from residual X band absorptivity above 51 km, with the assumption that the SO$_2$ abundance is constant within this range. Neither approach has thus far been used to place estimates on H$_2$SO$_4$ aerosol abundances. 

As we suggest in Figure \ref{fig:ex_mse_er}, dual X/Ka band radio occultations with future missions can likely provide improved accuracy over dual S/X band occultations due to the increased opacity of both H$_2$SO$_4$ aerosol and SO$_2$ gas. As with the S and X band measurements, this retrieval is both undetermined, since two measurements are used to determine 3 quantities), and ill-posed, as both absorbers have a roughly $\nu^2$ dependence in opacity between X and Ka band. It is therefore necessary to condition the retrieval using prior information. In the case of both species of absorbers, we are privy to few vertically resolved measurements, specifically the Pioneer Venus LCPS measurement of cloud aerosol mass \citep{Knollenberg1980} and the Vega descent probe SO$_2$ measurements \citep{Bertaux1996}. Of these, only the measured cloud aerosol mass is consistent with attempts to model the H$_2$SO$_4$ aerosol system, and the depletion mechanism of SO$_2$ within the clouds remain an area of active research (e.g. \cite{Rimmer2021}). The significant variations in SO$_2$ observed by the Vega landers have not been recreated in previous chemical or transport models, although ground-based observations suggest latitudinal variability that may be consistent with non-negligible sub-cloud abundance gradients \citep{Arney2014, Marcq2021}. If either the SO$_2$ or H$_2$SO$_4$ aerosol profile can be assumed known from the results of proximal in-situ measurements, uncertainties in the joint retrieval of the other with H$_2$SO$_4$ vapor will likely fall somewhere in between the cases illustrated in Figures \ref{fig:ex_mse_er} and \ref{fig:s18}. If the dual band measurements could sound below the cloud base,it would be possible to increase the accuracy of these retrievals by assigning the cloud-base SO$_2$ abundance. Unfortunately, the Ka band signal appears likely to become attenuation-limited near this altitude range \citep{Akins2020a}, and most measurements are unlikely to resolve the cloud base. 

The determination of cloud profiles from our 1D transport model is an extension of the information from the retrieved H$_2$SO$_4$ vapor profile. The variability of derived cloud profiles from the simulation ensemble is representative of the uncertainty in mean cloud structure. In the prior modeling efforts which inform this study, simulations are generally adjusted to achieve agreement with some observation, whether it be the LCPS measurement of aerosol mass or RO measurements of H$_2$SO$_4$ vapor. Our problem is the inverse, in that we are using such models to predict the abundances of cloud aerosols present in the measurement. We covered a range of possible simulation parameters in an attempt to mitigate the importance of model implementation details, such as model transport dimension or inclusion of more detailed cloud microphysics. It may be possible for a more realistic model of cloud aerosols to be brought to bear on this problem, but such an attempt would need to be based on high accuracy observations of a kind which do not exist at present. Generally, our simulated retrievals suggest that the profile variances assigned by the MCMC seeding method are useful if the underlying shape assumptions are valid. The autocorrelation time $\tau$ of the MCMC walkers (a metric of distribution convergence, \citep{Foreman-Mackey2013}) in our simulations is relatively long due to the multi-modal distribution of the resulting shape parameter fits. Our MCMC iteration number, however, is consistent with the suggested value of $50 \tau$, and the variances derived this way are representative of the distribution of possible profiles. 

Overall, the recovery of H$_2$SO$_4$ and SO$_2$ profiles simultaneously from dual X/Ka band RO measurements of Venus is an exceptional challenge, and as our simulations demonstrate, useful accuracy requires the incorporation of accurate prior information. Our proposed approach appears capable of conditioning the problem appropriately based on currently available information. If our assumptions are inaccurate, however, uncertainties in the retrieval of SO$_2$ and H$_2$SO$_4$ aerosol abundances could be significantly worse than those determined in our simulations. While the variance estimates provided using the proposed approach seem generally reliable, there are cases in which the true profile is not captured within the 1$\sigma$ estimate. Estimates of column abundances from the few retrieved profiles are encouraging, although not conclusive, with respect to the retrieval bias. The column accuracy of SO$_2$ in particular suggests that detection of time-variable enhancement associated with volcanism is likely achievable with this approach. While this is a challenging measurement, our simulations suggest that useful information can be obtained regarding the distributions of H$_2$SO$_4$ and SO$_2$ in clouds of Venus from future X and Ka band radio occultations.

\section{Conclusion} \label{sec:conclusion}
We have considered in detail the prospects for retrieving vertical profiles of H$_2$SO$_4$ (vapor and aerosol) and SO$_2$ abundance from dual X and Ka band radio occultation measurements of Venus which will be conducted by spacecraft missions in the near future. We first discussed the basis for measurement uncertainties that were used in this study, reviewed relevant models of atmospheric opacity derived from laboratory measurements, and derived formal uncertainty estimates for models of H$_2$SO$_4$ and SO$_2$ opacity. We then illustrated the ill-posedness of the retrieval problem and introduced a novel approach for seeding and regularizing maximum likelihood estimations of profile abundances. For the resulting retrievals we can estimate uncertainties of on the orders of 0.5 ppm for retrievals of H$_2$SO$_4$ vapor, 20 ppm for retrievals of SO$_2$, and 10 mg/m$^3$ for cloud aerosol bulk density. These uncertainty estimates are determined when the underlying assumptions informing the regularization are accurate, and we additionally discussed the implications of deviations from these assumptions on the retrieved abundances. From the retrieved column abundances, we surmise that the retrieval of SO$_2$ is more accurate than that of the cloud aerosol mass. Further ground-truth estimates from in situ measurements and more advanced atmospheric models can be used to further improve these results. We conclude that dual X/Ka band RO profiling of Venus' atmospheric sulfur species can be accomplished with sufficient accuracy ($<$ 50\% uncertainty in abundant regions) to provide useful insights into chemical and dynamical processes in the cloud-level atmosphere.

\section{Acknowledgements}
This work was funded by the JPL Research and Technology Development Fund. We would like to thank Janusz Oschlisniok and the VeRa team for providing processed X band radio occultation data used in  the model atmospheres. This work was carried out at the Jet Propulsion Laboratory, California Institute of Technology, under contract to the National Aeronautics and Space Administration

\bibliography{library.bib}

\begin{thebibliography}{}
\expandafter\ifx\csname natexlab\endcsname\relax\def\natexlab#1{#1}\fi
\providecommand{\url}[1]{\href{#1}{#1}}
\providecommand{\dodoi}[1]{doi:~\href{http://doi.org/#1}{\nolinkurl{#1}}}
\providecommand{\doeprint}[1]{\href{http://ascl.net/#1}{\nolinkurl{http://ascl.net/#1}}}
\providecommand{\doarXiv}[1]{\href{https://arxiv.org/abs/#1}{\nolinkurl{https://arxiv.org/abs/#1}}}

\bibitem[{Airey {et~al.}(2015)Airey, Mather, Pyle, Glaze, Ghail, \&
  Wilson}]{Airey2015}
Airey, M.~W., Mather, T.~A., Pyle, D.~M., {et~al.} 2015, Planetary and Space
  Science, 113-114, 33, \dodoi{10.1016/j.pss.2015.01.009}

\bibitem[{Akins \& Steffes(2019)}]{Akins2019}
Akins, A.~B., \& Steffes, P.~G. 2019, Icarus, 326, 18

\bibitem[{Akins \& Steffes(2020)}]{Akins2020a}
---. 2020, Icarus, 351, 113928, \dodoi{10.1016/j.icarus.2020.113928}

\bibitem[{Ando {et~al.}(2020)Ando, Imamura, Tellmann, Pätzold, Häusler,
  Sugimoto, Takagi, Sagawa, Limaye, Matsuda, Choudhary, \&
  Antonita}]{Ando2020a}
Ando, H., Imamura, T., Tellmann, S., {et~al.} 2020, Scientific Reports, 10, 1,
  \dodoi{10.1038/s41598-020-59278-8}

\bibitem[{Arney {et~al.}(2014)Arney, Meadows, Crisp, Schmidt, Bailey, \&
  Robinson}]{Arney2014}
Arney, G., Meadows, V., Crisp, D., {et~al.} 2014, Journal of Geophysical
  Research: Planets, 119, 1860, \dodoi{10.1002/2014JE004662}

\bibitem[{Barstow {et~al.}(2012)Barstow, Tsang, Wilson, Irwin, Taylor,
  McGouldrick, Drossart, Piccioni, \& Tellmann}]{Barstow2012}
Barstow, J.~K., Tsang, C.~C., Wilson, C.~F., {et~al.} 2012, Icarus, 217, 542,
  \dodoi{10.1016/j.icarus.2011.05.018}

\bibitem[{Bellotti \& Steffes(2015)}]{Bellotti2015}
Bellotti, A., \& Steffes, P.~G. 2015, Icarus, 254, 24,
  \dodoi{10.1016/j.icarus.2015.03.028}

\bibitem[{Bertaux {et~al.}(1996)Bertaux, Widemann, Hauchecorne, Moroz, \&
  Ekonomov}]{Bertaux1996}
Bertaux, J.-L., Widemann, T., Hauchecorne, A., Moroz, V.~I., \& Ekonomov, A.~P.
  1996, Journal of Geophysical Research-Planets, 101, 12709,
  \dodoi{10.1029/96JE00466}

\bibitem[{Bierson \& Zhang(2020)}]{Bierson2020}
Bierson, C.~J., \& Zhang, X. 2020, Journal of Geophysical Research: Planets,
  125, 1, \dodoi{10.1029/2019JE006159}

\bibitem[{Bullock \& Grinspoon(2001)}]{Bullock2001}
Bullock, M.~A., \& Grinspoon, D.~H. 2001, Icarus, 150, 19,
  \dodoi{10.1006/icar.2000.6570}

\bibitem[{Eshleman(1973)}]{Eshleman1973}
Eshleman, V.~R. 1973, Planetary and Space Science, 21, 1521.
\newblock
  \url{http://ac.els-cdn.com/0032063373900597/1-s2.0-0032063373900597-main.pdf?_tid=c65cb660-658b-11e7-bbeb-00000aab0f27&acdnat=1499703846_ff15371f131fab9d32cead6bf29c9f9b}

\bibitem[{Fahd \& Steffes(1991)}]{Fahd1991}
Fahd, A.~K., \& Steffes, P.~G. 1991, Journal of Geophysical Research: Planets,
  96, 17471

\bibitem[{Fahd \& Steffes(1992)}]{Fahd1992}
---. 1992, Icarus, 97, 200, \dodoi{10.1016/0019-1035(92)90128-T}

\bibitem[{Fjeldbo {et~al.}(1971)Fjeldbo, J.~Kliore, \& Eshleman}]{Fjeldbo1971}
Fjeldbo, G., J.~Kliore, A., \& Eshleman, V.~R. 1971, The Astronomical Journal,
  76

\bibitem[{Foreman-Mackey {et~al.}(2013)Foreman-Mackey, Hogg, Lang, \&
  Goodman}]{Foreman-Mackey2013}
Foreman-Mackey, D., Hogg, D.~W., Lang, D., \& Goodman, J. 2013, Publications of
  the Astronomical Society of the Pacific, 125, 306, \dodoi{10.1086/670067}

\bibitem[{Glaze(1999)}]{Glaze1999}
Glaze, L.~S. 1999, Journal of Geophysical Research: Planets, 104, 18899,
  \dodoi{10.1029/1998JE000619}

\bibitem[{Hashimoto \& Abe(2001)}]{Hashimoto2001}
Hashimoto, G.~L., \& Abe, Y. 2001, Journal of Geophysical Research: Planets,
  106, 14675, \dodoi{10.1029/2000JE001266}

\bibitem[{Haus {et~al.}(2013)Haus, Kappel, \& Arnold}]{Haus2013}
Haus, R., Kappel, D., \& Arnold, G. 2013, Planetary and Space Science, 89, 77,
  \dodoi{10.1016/j.pss.2013.09.020}

\bibitem[{Haüsler {et~al.}(2006)Haüsler, Pätzold, Tyler, Simpson, Bird,
  Dehant, Barriot, Eidel, Mattei, Remus, Selle, Tellmann, \&
  Imamura}]{Hausler2006a}
Haüsler, B., Pätzold, M., Tyler, G.~L., {et~al.} 2006, Planetary and Space
  Science, 54, 1315, \dodoi{10.1016/j.pss.2006.04.032}

\bibitem[{Ho {et~al.}(1966)Ho, Kaufman, \& Thaddeus}]{Ho1966}
Ho, W., Kaufman, I.~A., \& Thaddeus, P. 1966, Journal of Geophysical Research,
  71, 5091, \dodoi{10.1029/JZ071i021p05091}

\bibitem[{Imamura \& Hashimoto(1998)}]{Imamura1998}
Imamura, T., \& Hashimoto, G.~L. 1998, Journal of Geophysical Research, 103,
  31349, \dodoi{10.1029/1998JE900010}

\bibitem[{Imamura \& Hashimoto(2001)}]{Imamura2001}
---. 2001, Journal of the Atmospheric Sciences, 58, 3597,
  \dodoi{10.1175/1520-0469(2001)058<3597:MOVCIR>2.0.CO;2}

\bibitem[{Imamura {et~al.}(2017)Imamura, Ando, Tellmann, Pätzold, Häusler,
  Yamazaki, Sato, Noguchi, Futaana, Oschlisniok, Limaye, Choudhary, Murata,
  Takeuchi, Hirose, Ichikawa, Toda, Tomiki, Abe, ichi Yamamoto, Noda, Iwata,
  ya~Murakami, Satoh, Fukuhara, Ogohara, ichiro Sugiyama, Kashimura, Ohtsuki,
  Takagi, Yamamoto, Hirata, Hashimoto, Yamada, Suzuki, Ishii, Hayashiyama, Lee,
  \& Nakamura}]{Imamura2017}
Imamura, T., Ando, H., Tellmann, S., {et~al.} 2017, Earth, Planets and Space,
  69, 137, \dodoi{10.1186/s40623-017-0722-3}

\bibitem[{James {et~al.}(1997)James, Toon, \& Schubert}]{James1997}
James, E.~P., Toon, O.~B., \& Schubert, G. 1997, Icarus, 129, 147,
  \dodoi{10.1006/icar.1997.5763}

\bibitem[{Jenkins(1992)}]{Jenkins1992}
Jenkins, J.~M. 1992, Variations on the 13 cm Opacity below the Main Cloud Layer
  in the Atmosphere of Venus Inferred from Pioneer-Venus Radio Occultation
  Studies 1978-1987

\bibitem[{Jenkins {et~al.}(2002)Jenkins, Kolodner, Butler, Suleiman, \&
  Steffes}]{Jenkins2002}
Jenkins, J.~M., Kolodner, M.~A., Butler, B.~J., Suleiman, S.~H., \& Steffes,
  P.~G. 2002, Icarus, 158, 312,
  \dodoi{http://dx.doi.org/10.1006/icar.2002.6894}

\bibitem[{Jenkins \& Steffes(1991)}]{Jenkins1991}
Jenkins, J.~M., \& Steffes, P.~G. 1991, Icarus, 90, 129,
  \dodoi{10.1016/0019-1035(91)90075-5}

\bibitem[{Jenkins {et~al.}(1994)Jenkins, Steffes, Hinson, Twicken, \&
  Tyler}]{Jenkins1994}
Jenkins, J.~M., Steffes, P.~G., Hinson, D.~P., Twicken, J.~D., \& Tyler, G.~L.
  1994, Icarus, 110, 79, \dodoi{10.1006/icar.1994.1108}

\bibitem[{Knollenberg \& Hunten(1980)}]{Knollenberg1980}
Knollenberg, R.~G., \& Hunten, D.~M. 1980, Journal of Geophysical Research, 85,
  8039, \dodoi{10.1029/JA085iA13p08039}

\bibitem[{Kolodner \& Steffes(1998)}]{Kolodner1998a}
Kolodner, M.~A., \& Steffes, P.~G. 1998, Icarus, 132, 151,
  \dodoi{10.1006/icar.1997.5887}

\bibitem[{Krasnopolsky(2012)}]{Krasnopolsky2012}
Krasnopolsky, V.~A. 2012, Icarus, 218, 230,
  \dodoi{10.1016/j.icarus.2011.11.012}

\bibitem[{Krasnopolsky(2015)}]{Krasnopolsky2015}
---. 2015, Icarus, 252, 327, \dodoi{10.1016/j.icarus.2015.01.024}

\bibitem[{Limaye {et~al.}(2018)Limaye, Grassi, Mahieux, Migliorini, Tellmann,
  \& Titov}]{Limaye2018a}
Limaye, S.~S., Grassi, D., Mahieux, A., {et~al.} 2018, Space Science Reviews,
  214, 102, \dodoi{10.1007/s11214-018-0525-2}

\bibitem[{Lipa \& Tyler(1979)}]{Lipa1979}
Lipa, B., \& Tyler, G.~L. 1979, Icarus, 39, 192,
  \dodoi{10.1016/0019-1035(79)90163-5}

\bibitem[{Marcq {et~al.}(2021)Marcq, Amine, Duquesnoy, \& Bézard}]{Marcq2021}
Marcq, E., Amine, I., Duquesnoy, M., \& Bézard, B. 2021, Astronomy and
  Astrophysics, 648, L8, \dodoi{10.1051/0004-6361/202140837}

\bibitem[{Marcq {et~al.}(2013)Marcq, Bertaux, Montmessin, \&
  Belyaev}]{Marcq2013}
Marcq, E., Bertaux, J.-L., Montmessin, F., \& Belyaev, D.~A. 2013, Nature
  Geoscience, 6, 25.
\newblock \url{https://www.nature.com/ngeo/journal/v6/n1/pdf/ngeo1650.pdf}

\bibitem[{McGouldrick \& Toon(2007)}]{McGouldrick2007}
McGouldrick, K., \& Toon, O.~B. 2007, Icarus, 191, 1,
  \dodoi{10.1016/j.icarus.2007.04.007}

\bibitem[{Oschlisniok(2020)}]{Oschlisniok2020}
Oschlisniok, J. 2020, Transport of sulfuric acid in the atmosphere of Venus
  studied on the basis of radio signal attenuation effects observed in the
  Venus Express Radio Science Experiment VeRa

\bibitem[{Oschlisniok {et~al.}(2021)Oschlisniok, Häusler, Pätzold, Tellmann,
  Bird, Peter, \& Andert}]{Oschlisniok2021}
Oschlisniok, J., Häusler, B., Pätzold, M., {et~al.} 2021, Icarus, 362,
  114405, \dodoi{10.1016/j.icarus.2021.114405}

\bibitem[{Oschlisniok {et~al.}(2012)Oschlisniok, Häusler, Pätzold, Tyler,
  Bird, Tellmann, Remus, \& Andert}]{Oschlisniok2012}
---. 2012, Icarus, 221, 940, \dodoi{10.1016/j.icarus.2012.09.029}

\bibitem[{Pätzold {et~al.}(2007)Pätzold, Häusler, Bird, Tellmann, Mattei,
  Asmar, Dehant, Eidel, Imamura, Simpson, \& Tyler}]{Patzold2007}
Pätzold, M., Häusler, B., Bird, M.~K., {et~al.} 2007, Nature, 450, 657,
  \dodoi{10.1038/nature06239}

\bibitem[{Rimmer {et~al.}(2021)Rimmer, Jordan, Constantinou, Woitke, Shorttle,
  Hobbs, \& Paschodimas}]{Rimmer2021}
Rimmer, P.~B., Jordan, S., Constantinou, T., {et~al.} 2021, The Planetary
  Science Journal, 2, 133, \dodoi{10.3847/psj/ac0156}

\bibitem[{Rodgers(2000)}]{Rodgers2000}
Rodgers, C.~D. 2000, Inverse Methods for Atmospheric Sounding: Theory and
  Practice (World Scientific)

\bibitem[{Spiegelman \& Katz(2006)}]{Spiegelman2006}
Spiegelman, M., \& Katz, R.~F. 2006, Geochemistry, Geophysics, Geosystems, 7,
  1, \dodoi{10.1029/2005GC001073}

\bibitem[{Steffes \& Eshleman(1982)}]{Steffes1982}
Steffes, P.~G., \& Eshleman, V.~R. 1982, Icarus, 51, 322,
  \dodoi{10.1016/0019-1035(82)90087-2}

\bibitem[{Steffes {et~al.}(2015)Steffes, Shahan, Barisich, \&
  Bellotti}]{Steffes2015}
Steffes, P.~G., Shahan, P., Barisich, G.~C., \& Bellotti, A. 2015, Icarus, 245,
  153, \dodoi{10.1016/j.icarus.2014.09.012}

\bibitem[{Suleiman {et~al.}(1996)Suleiman, Kolodner, \& Steffes}]{Suleiman1996}
Suleiman, S.~H., Kolodner, M.~A., \& Steffes, P.~G. 1996, Journal of
  Geophysical Research: Planets, 101, 4623

\bibitem[{Sánchez-Lavega {et~al.}(2017)Sánchez-Lavega, Lebonnois, Imamura,
  Read, \& Luz}]{Sanchez-Lavega2017}
Sánchez-Lavega, A., Lebonnois, S., Imamura, T., Read, P., \& Luz, D. 2017,
  Space Science Reviews, 212, 1541, \dodoi{10.1007/s11214-017-0389-x}

\bibitem[{Team(2021)}]{Envision2021}
Team, E. S.~S. 2021, {EnVision Assessment Study Report (Yellow Book)}, Tech.
  rep., European Space Agency

\bibitem[{Tellmann {et~al.}(2012)Tellmann, Häusler, Hinson, Tyler, Andert,
  Bird, Imamura, Pätzold, \& Remus}]{Tellmann2012}
Tellmann, S., Häusler, B., Hinson, D.~P., {et~al.} 2012, Icarus, 221, 471,
  \dodoi{10.1016/j.icarus.2012.08.023}

\bibitem[{Vandaele {et~al.}(2017)Vandaele, Korablev, Belyaev, Chamberlain,
  Evdokimova, Encrenaz, Esposito, Jessup, Lefèvre, Limaye, Mahieux, Marcq,
  Mills, Montmessin, Parkinson, Robert, Roman, Sandor, Stolzenbach, Wilson, \&
  Wilquet}]{Vandaele2017a}
Vandaele, A.~C., Korablev, O., Belyaev, D.~A., {et~al.} 2017, Icarus, 295, 1,
  \dodoi{10.1016/j.icarus.2017.05.001}

\bibitem[{Virtanen {et~al.}(2020)Virtanen, Gommers, Oliphant, Haberland, Reddy,
  Cournapeau, Burovski, Peterson, Weckesser, Bright, van~der Walt, Brett,
  Wilson, Millman, Mayorov, Nelson, Jones, Kern, Larson, Carey, Polat, Feng,
  Moore, VanderPlas, Laxalde, Perktold, Cimrman, Henriksen, Quintero, Harris,
  Archibald, Ribeiro, Pedregosa, van Mulbregt, \&
  Contributors}]{2020SciPy-NMeth}
Virtanen, P., Gommers, R., Oliphant, T.~E., {et~al.} 2020, Nature Methods, 17,
  261, \dodoi{10.1038/s41592-019-0686-2}

\bibitem[{Withers {et~al.}(2020)Withers, Hensley, Vogt, \&
  Hermann}]{Withers2020}
Withers, P., Hensley, K., Vogt, M.~F., \& Hermann, J. 2020, The Planetary
  Science Journal, 1, 79, \dodoi{10.3847/psj/abc476}

\bibitem[{Zhang {et~al.}(2012)Zhang, Liang, Mills, Belyaev, \&
  Yung}]{Zhang2012}
Zhang, X., Liang, M.~C., Mills, F.~P., Belyaev, D.~A., \& Yung, Y.~L. 2012,
  Icarus, 217, 714, \dodoi{10.1016/j.icarus.2011.06.016}

\end{thebibliography}
\bibliographystyle{aasjournal}

\end{document}